\documentclass[11pt,a4paper]{article}
\usepackage{jcappub}
\pdfoutput=1
\usepackage{epstopdf}
\usepackage{slashed}
\usepackage{subfig}
\usepackage {amsfonts}
\usepackage{comment}

\newcommand{\BigO}[1]{\ensuremath{\operatorname{\mathcal{O}}\left(#1\right)}}
 
 \newcommand{\EWbasin}{\Delta_{EW}} 
 \newcommand{\UVbasin}{\Delta_{UV}} 
 \newcommand{\minEW}{h_{EW}} 
 \newcommand{\minUV}{h_{UV}} 

\begin{document} 

\title{Domain walls in the extensions of the Standard Model}

\author[a]{Tomasz~Krajewski,}
\emailAdd{Tomasz.Krajewski@fuw.edu.pl}
\author[a]{Zygmunt~Lalak,}
\emailAdd{Zygmunt.Lalak@fuw.edu.pl}
\author[b]{Marek~Lewicki,}
\emailAdd{Marek.Lewicki@adelaide.edu.au}
\author[a]{Pawe\l~Olszewski}
\emailAdd{Pawel.Olszewski@fuw.edu.pl}
\affiliation[a]{Institute of Theoretical Physics, Faculty of Physics, University of Warsaw,\\
ul. Pasteura 5, Warsaw, Poland}
\affiliation[b]{ARC Centre of Excellence for Particle Physics at the Terascale (CoEPP) \& CSSM,\\
Department of Physics, University of Adelaide, South Australia 5005, Australia}

\abstract{
Our main interest is the evolution of domain walls of the Higgs field in the early Universe. The aim of this paper is to understand how dynamics of Higgs domain walls could be influenced by yet unknown interactions from beyond the Standard Model. We assume that the Standard Model is valid up to certain, high, energy scale $\Lambda$ and use the framework of the effective field theory to describe physics below that scale. Performing numerical simulations with different values of the scale $\Lambda$ we are able to extend our previous analysis \cite{Krajewski:2016vbr} and determine its range of validity.

We study domain walls interpolating between the physical electroweak vacuum and the vacuum appearing at very high field strengths. These domain walls could be formed from non-homogeneous configurations of the Higgs field produced by quantum fluctuations during inflation or thermal fluctuations during reheating.

Our numerical simulations show that evolution of Higgs domain walls is rather insensitive to interactions beyond the Standard Model as long as masses of new particles are grater than $10^{12}\; \textrm{GeV}$.
For lower values of $\Lambda$ the RG improved effective potential is strongly modified at field strengths crucial to the evolution of domain walls. For instance its minima become degenerate for $\Lambda$ around $10^{11}\; \textrm{GeV}$.
We find that even in the case when the minima of the potential are nearly degenerate Higgs domain walls decayed shortly after their formation for generic initial conditions.
On the other hand, in simulations with specifically chosen initial conditions Higgs domain walls can live longer and enter the scaling regime.

We also determine the energy spectrum of gravitational waves produced by decaying domain walls of the Higgs field. For generic initial field configurations the amplitude of the signal is too small to be observed in present and planned detectors.
}


\notoc

\maketitle

\section{Introduction}
The precise measurement of the Higgs boson mass at the Large Hadron Collider \cite{Aad:2012tfa,Chatrchyan:2012ufa} enabled study of the scalar effective potential in the Standard Model up to superplanckian field values.
The potential at field strengths higher than approximately $10^{11}$ GeV drops below its value at the electroweak vacuum and develops a new minimum at superplanckian field strengths. This opens the possibility of tunneling from the physical electroweak vacuum to the deeper minimum at very large Higgs field value. This effect could render the electroweak vacuum unstable and has been widely discussed in the literature \cite{Andreassen:2017rzq,Salvio:2016mvj,Lalak:2014qua,Andreassen:2014gha,Buttazzo:2013uya,Degrassi:2012ry,Ellis:2009tp,Espinosa:2007qp,Casas:2000mn,Casas:1996aq,Casas:1994qy,Sher:1988mj}. Metastability of the electroweak vacuum for present central values of measured parameters of the SM has been shown under the assumption that the SM is valid up to the Planck scale. However, the problem of stability of the electroweak vacuum can be modified, especially if particles with masses of the order of $10^{11}$ GeV or lower, not described by the SM, exist in nature. This scenario has been investigated in \cite{Lalak:2014qua} in the framework of the effective field theory. Furthermore, even if the electroweak vacuum is metastable, existence of the second deeper minimum poses severe problems for cosmological models~\cite{Kobakhidze:2014xda,East:2016anr,Joti:2017fwe}.
 
One of the most promising solutions to problems of the standard cosmology is the occurrence of an inflationary epoch. The analysis based on the Fokker-Planck equation \cite{Espinosa:2007qp,Hook:2014uia,Espinosa:2015qea,East:2016anr,Darme:2017wvu} generally predicts that the Higgs field distribution after inflation is, in the first approximation, Gaussian with the standard deviation of the order of 
\begin{equation}
\sigma_I \sim \frac{\sqrt{N} H_I}{2 \pi},
\end{equation}
where $H_I$ is the value of the Hubble constant during the inflation and $N$ is the number of e-folds which for experimentally viable models exceeds value of 50--60.

We will investigate the case in which Hubble constant during inflation was of the order of the position of the local maximum $h_{max}$ separating the two minima in the Higgs potential.
Thus, the generated field fluctuations would be large enough to create a network of domain walls which interpolate between areas of the Universe occupied by the Higgs field lying in different minima of the potential.

We perform an analysis of the evolution of these structures using numerical lattice simulations based on the Press, Ryden, Spergel (PRS) algorithm \cite{Press:1989yh}.

We will refer to the basin of attraction of the electroweak vacuum, that is the range of fields whose absolute values are less than $h_{max}$, as $\EWbasin$. Analogously $\UVbasin$ will denote the range of fields of an absolute value greater than $h_{max}$.
\begin{equation}
\Delta_{EW(UV)} = \left\{h:\;\left|h\right| <(>)\; h_{max} \right\}
\end{equation} 
Let us also adopt labels for the exact field values that minimize the potential: $\minEW$ for the electroweak vacuum and $\minUV$ for the high energy vacuum.

Domain walls, interpolating between non-degenerate minima are known to be generically unstable \cite{Lalak:2007rs}. As we have shown in our previous article \cite{Krajewski:2016vbr}, assuming validity of the SM up to the Planck scale, networks of Higgs domain walls decayed in conformal time of the order of $10^{-9}\ \textrm{GeV}^{-1}$. Initial configurations of the Higgs field with at least 49 \% of the expectation values belonging to $\UVbasin$ decayed to the superplanckian minimum. Models of the early Universe predicting such configurations of the Higgs field are excluded by experiments. On the other hand for networks initialised with the higher fraction of lattice sites belonging to $\EWbasin$, the final state of their evolution was the electroweak vacuum.
 Moreover, the short decay time of these networks guarantee consistency with present experimental data.

In this article we investigate the possibility that there exist new, yet unknown physical phenomena which are not described by the SM. We parametrize the influence of these new interactions on the physics of the Higgs boson in the framework of the effective field theory (EFT). To that end we introduce a nonrenormalizable operator $h^6$ suppressed by the scale $\Lambda$ in the Lagrangian density of the SM.
Performing simulations with different values of the initial standard deviation of the field strength $\sigma_I$ and different values of the suppression scale $\Lambda$, we are able to determine the influence of this new interaction on evolution of Higgs domain walls. We find that a naive expectation that new physics will not modify dynamics of Higgs domain walls if the suppression scale will be much higher then the position of the local maximum of the effective potential $h_{max}$ is accurate. Moreover we have considered lower suppression scales leading to a~reduced difference between values of the effective potential in the two minima. The scenario with nearly degenerate minima is interesting, because it allows metastable networks of domain walls which would lead to a reach variety of phenomenological implications.

During the process of decay of domain walls, the energy of the field is transferred to other degrees of freedom and both SM particles and gravitational waves (GWs) can be produced. The next point in considering low suppression scales producing nearly degenerate minima is the possibility to increase the energy density of produced GWs. In our previous work we estimated the energy spectrum of GWs produced by decaying domain walls of the Higgs field in the SM valid up to the Planck scale. The obtained present energy density of GWs was orders of magnitude smaller than the sensitivity of present and planned detectors of GWs and so we excluded the possibility of their detection in the near future.
However, our previous results do not apply to the case of metastable domain walls associated with nearly degenerate minima of the potential. Authors of \cite{Kitajima:2015nla} claim, basing on a~semi-analytical approximation \cite{Hiramatsu:2013qaa}, that GWs emitted from decaying Higgs domain walls in the case of nearly degenerate minima can be strong enough to be measured in planned detectors. In our numerical simulations we investigate evolution of Higgs domain walls with the potential with nearly degenerate minima. Unfortunately, we were not able to study the scenario proposed in \cite{Kitajima:2015nla} with very long decay time of the network of domain walls using lattice simulations due to their limited dynamical range.

Our simulations show that Higgs domain walls decay in conformal time shorter than $10^{-8}\ \textrm{GeV}^{-1}$ after their formation if generic initial configurations of the Higgs field are assumed. 
We were however, able to simulate networks with decay times up to five times longer by setting very specific initial conditions (to compensate for the asymmetry of the potential around the local maximum) as well as a~fine-tuned coupling constant for the $h^6$ interaction to obtain nearly degenerate minima. These examples show the possibility of formation of metastable networks of domain walls when the minima of the potential are nearly degenerate. Moreover, we prove that these long-lived networks of domain walls evolve in the scaling regime, satisfying assumptions of the semi-analytical method used in \cite{Kitajima:2015nla}. However, the formation of networks evolving in the scaling regime requires specifically chosen initial conditions in addition to assumptions used in \cite{Kitajima:2015nla}. Unfortunately, removing unphysical, short wave-lengths modes in the initial configuration of the field what is necessary for computing GWs spectrum in finite lattice simulations, we also spoil the specific choice of the initial configuration required for the networks longevity. 

The paper is organised as follows. In section \ref{SM_potential} we briefly present the form of the RG improved effective potential of the SM in the presence of the nonrenormalizable operator $h^6$. The influence of the modified potential on properties of Higgs domain walls is analysed in section \ref{wall}. We discuss initial conditions for our simulations in section~\ref{initial}. Next, in section~\ref{decay_time} we discuss dependence of the decay time of networks of domain walls on the suppression scale $\Lambda$. Subsection \ref{metastability} is devoted to an~investigation of the possible meta-stability of networks of domain walls.
Finally, we calculate the energy spectrum of GWs for the case of nearly degenerated minima in section \ref{spectrum}. Our results are summarized in section \ref{summary}.

\section{RG improved SM effective potential with $h^6$ operator\label{SM_potential}}
In our previous studies \cite{Krajewski:2016vbr} we have investigated evolution of domain walls of the SM Higgs field. The aim of this study is to understand the way in which our previous results can be modified by new, not yet discovered interactions of the Higgs boson. 

The starting point of our considerations is the zero temperature scalar effective potential in the SM which takes the form \cite{Buttazzo:2013uya}:
\begin{gather}
V_{\text{SM}}(h;\mu) = \frac{1}{2} m^2(\mu) h^2 + \frac{1}{4} \lambda_{\text{eff}}(h; \mu) h^4, \\
\lambda_{\text{eff}}(h;\mu) = \lambda(\mu) + \lambda_{\text{eff}}^{(1)}(h;\mu) + \lambda_{\text{eff}}^{(2)}(h;\mu)\;,
\end{gather}
which depends implicitly on the running couplings $m^2$, $\lambda$, $g_1$, $g_2$, $g_3$, $y_{top}$ (only the largest Yukawa coupling of the third family is included). Explicit $n$-loop corrections to the quartic term are gathered in $\lambda_{\text{eff}}^{(n)}(h;\mu)$, where the field variable $h$ appears only as $h/\mu$. Hence, the substitution $\mu = h$ allows one to resum large logarithms by solving renormalization group equations (RGEs) to appropriately high scales. We use three-loop RGEs and boundary conditions together with $\lambda_{\text{eff}}$ determined up to two-loops. This leaves us with the effective potential as a function of a single variable,
\begin{equation}
\widetilde{V}_{\text{SM}}(h) \equiv V_{\text{SM}}(h; h)\;.
\end{equation}
This potential is famously unstable and exhibits a~double near-criticality in its dependence on the masses of the Higgs boson and the top quark, discussed for example in \cite{Buttazzo:2013uya}.

We consider modifications of $\widetilde{V}_{\text{SM}}(h)$ that might stem from hypothetical existence of heavier particles, not included in the determination of the potential in the SM alone.

The EFT provides a computationally convenient framework for treating corrections from high energy physics to observables at lower energies via integrating out heavy states. This paradigm states that quantum corrections to scattering amplitudes of light modes at low energies coming from heavy states (whose masses are much higher relative to four-momenta of the light modes) can be reconstructed (in a perturbative expansion) by inclusion of nonrenormalizable operators with properly adjusted coupling constants into the Lagrangian density of the light fields. Due to dimensional arguments, the coupling constants of the nonrenormalizable operators have to be suppressed by inverse powers of certain energy scale $\Lambda$ which is naturally of the order of masses of heavy states. Reversing the reasoning, the EFT approach can be used to parametrise, in a model independent way, the yet unknown high energy physics when treating values of coupling constants of nonrenormalizable operators as free parameters.

Basis of nonrenormalizable operators up to dimension 6 for SM fields was determined \cite{Buchmuller:1985jz,Grzadkowski:2010es} in the past. We will concentrate our analysis on the operator $h^6$ which influences the effective scalar potential the most since it is the only dimension 6 interaction term that simply enters the potential at the tree-level. Thus we choose to examine potential of the form:
\begin{equation}
V_\text{SM}^\Lambda(h) = \widetilde{V}_\text{SM}(h) + \frac{\lambda_6}{6!}\frac{h^6}{M^2} = \widetilde{V}_\text{SM}(h) + \frac{1}{6!}\frac{h^6}{\Lambda^2}\;, \label{VLambda}
\end{equation}
where $M$ is the energy scale used in the numerical integration of the RGEs \cite{Lalak:2014qua}. We also define the scale $\Lambda$ which unambiguously parametrises the coupling constant of the nonrenormalizable operator $h^6$.
We simply use function $V_\text{SM}^\Lambda$ as a~one-parameter modification of $\widetilde{V}_{\text{SM}}$ with the robust characteristic that a significant deviation is switched on at a particular scale, $h\sim\Lambda$. 
It is highly debatable, we realize, whether value of the coupling $\lambda_6$ hypothetically determined in low energy expansion of nonrenormalizable corrections, valid for $h \ll \Lambda$, can be meaningfully used in \eqref{VLambda} to encode new physics in its entirety when $h$ is not much smaller than $\Lambda$. Following other authors \cite{Grzadkowski:2001vb, Burgess:2001tj, Branchina:2014rva}, we will still use $\Lambda$ as a~free parameter which lets us smoothly interpolate between the scalar potential of the pure SM and the Higgs potential with nearly degenerate minima. 

Scenario in which new interactions lift the value of the effective potential at the high field strength minimum $V^\Lambda_\text{SM}(\minUV)$ up to the value at the electroweak one $V^\Lambda_\text{SM}(\minEW)$ is supposed to have very interesting phenomenology. The general prediction states that networks of cosmological domain walls corresponding to a~potential with degenerate minima should be metastable. Due to their long decay times, and consequently large abundance in late cosmological epochs, such networks might imprint sizable distortions in cosmological observables.

In the case of Higgs domain walls the most promising observable is the energy density spectrum of GWs produced by their decay. Previously we argued \cite{Krajewski:2016vbr} that if the SM is valid up to the Planck scale, GWs produced by Higgs domain walls are orders of magnitude to weak to be detected in present or planned experiments. The main reason of such a~small amplitude of the signal is the short life-time of Higgs domain walls. Generally, the energy density of a network of domain walls decreases slower then the energy density of both: radiation and dust. Thus, the network's contribution to the total energy density of the Universe grows in cosmic time. As a~result, the longer the network lives, the higher abundance of GWs $\Omega_{GW}$ it produces. 

The energy spectrum of GWs emitted by Higgs domain walls with nearly degenerate minima of the scalar potential has been previously studied in \cite{Kitajima:2015nla} with the semi-analytical approach. This method is reliable if domain walls evolve in the so called scaling regime, during which the number of walls in the Hubble horizon is approximately preserved in time. The authors of \cite{Kitajima:2015nla} claim that GWs emitted from Higgs domain walls can be detected by future detectors if the difference between values of the Higgs potential in two minima is very small.

However, a small difference between values of the Higgs potential at minima requires a~severe fine-tuning of the value of the coupling constant $\lambda_6$. The condition for nearly degenerate minima can be written down as:
\begin{equation}
\begin{split}
&(1 + \varepsilon) V^\Lambda_{\text{SM}} (\minEW) =\\
& = (1 + \varepsilon) \left[\frac{1}{2} m^2(\minEW) \minEW^2 + \frac{1}{4} \lambda_{\text{eff}}(\minEW; \minEW) \minEW^4 +\frac{1}{6!} {\lambda_6}(\minEW; \minEW) \frac{\minEW^6}{M^2} + L\right] =\\
& = \left[\frac{1}{2} m^2(\minUV) \minUV^2 + \frac{1}{4} \lambda_{\text{eff}}(\minUV; \minUV) \minUV^4 +\frac{1}{6!} {\lambda_6}(\minUV; \minUV) \frac{\minUV^6}{M^2}\right] + L\\
& = V^\Lambda_{\text{SM}} (\minUV),
\end{split}
\end{equation}
where $\varepsilon$ is a~small number that parametrizes the degeneracy of both minima and $L$ is a~constant whose value is determined by the renormalization of the cosmological constant.
After dividing both sides by $ \minEW^4$ and suppressing the constant $L$ we get:
\begin{equation}
\begin{split}
&\frac{1}{2} \frac{m^2(\minEW)}{\minEW^2} + \frac{1}{4} \lambda_{\text{eff}}(\minEW; \minEW) +\frac{1}{6!} {\lambda_6}(\minEW; \minEW) \frac{\minEW^2}{M^2} =\\
& = (1 + \varepsilon)^{-1} \left(\frac{\minUV}{\minEW}\right)^4 \left[\frac{1}{2} \frac{m^2(\minUV)}{\minEW^2} \left(\frac{\minEW}{\minUV}\right)^2 + \frac{1}{4} \lambda_{\text{eff}}(\minUV; \minUV) +\frac{1}{6!} {\lambda_6}(\minUV; \minUV) \frac{\minUV^2}{M^2} \right]
\end{split}
\end{equation}

At this point the first conclusion from numerical studies of the RG improved potential of the SM with the operator $h^6$ is that nearly degenerate minima are obtained for suppression scales of the order of $\Lambda \sim 10^{11}\; \textrm{GeV}$, so ${\lambda_6}(\minEW; \minEW) \frac{\minEW^2}{M^2} \sim \BigO{10^{-9}}$. Moreover the experimental data determines $\frac{m^2(\minEW)}{\minEW^2} \sim \BigO{1}$ and $\lambda_{\text{eff}}(\minEW; \minEW) \sim \BigO{0.1}$. Thus $\frac{V^\Lambda_{\text{SM}} (\minEW) - L}{\minEW^4} \sim \BigO{0.1}$.

The numerical study of the RG improved SM effective potential reveals that for nearly degenerate minima $\minUV \sim 10^{10} \textrm{GeV}$. This leads to the conclusion that a~natural scale for $V^\Lambda_{\text{SM}} (\minUV) - L$ is $ \left(\frac{\minUV}{\minEW}\right)^4 \sim 10^{32} \; \textrm{GeV}$, so in order to produce nearly degenerate minima the $\lambda_6$ coupling constant needs to be fine-tuned in order to get the value of the factor $\frac{1}{2} \frac{m^2(\minUV)}{\minEW^2} \left(\frac{\minEW}{\minUV}\right)^2 + \frac{1}{4} \lambda_{\text{eff}}(\minUV; \minUV) +\frac{1}{6!} {\lambda_6}(\minUV; \minUV) \frac{\minUV^2}{M^2}$ down to $\BigO{10^{-32}}$.

\begin{figure}[t]
\centering
\includegraphics[width= 0.7\textwidth]{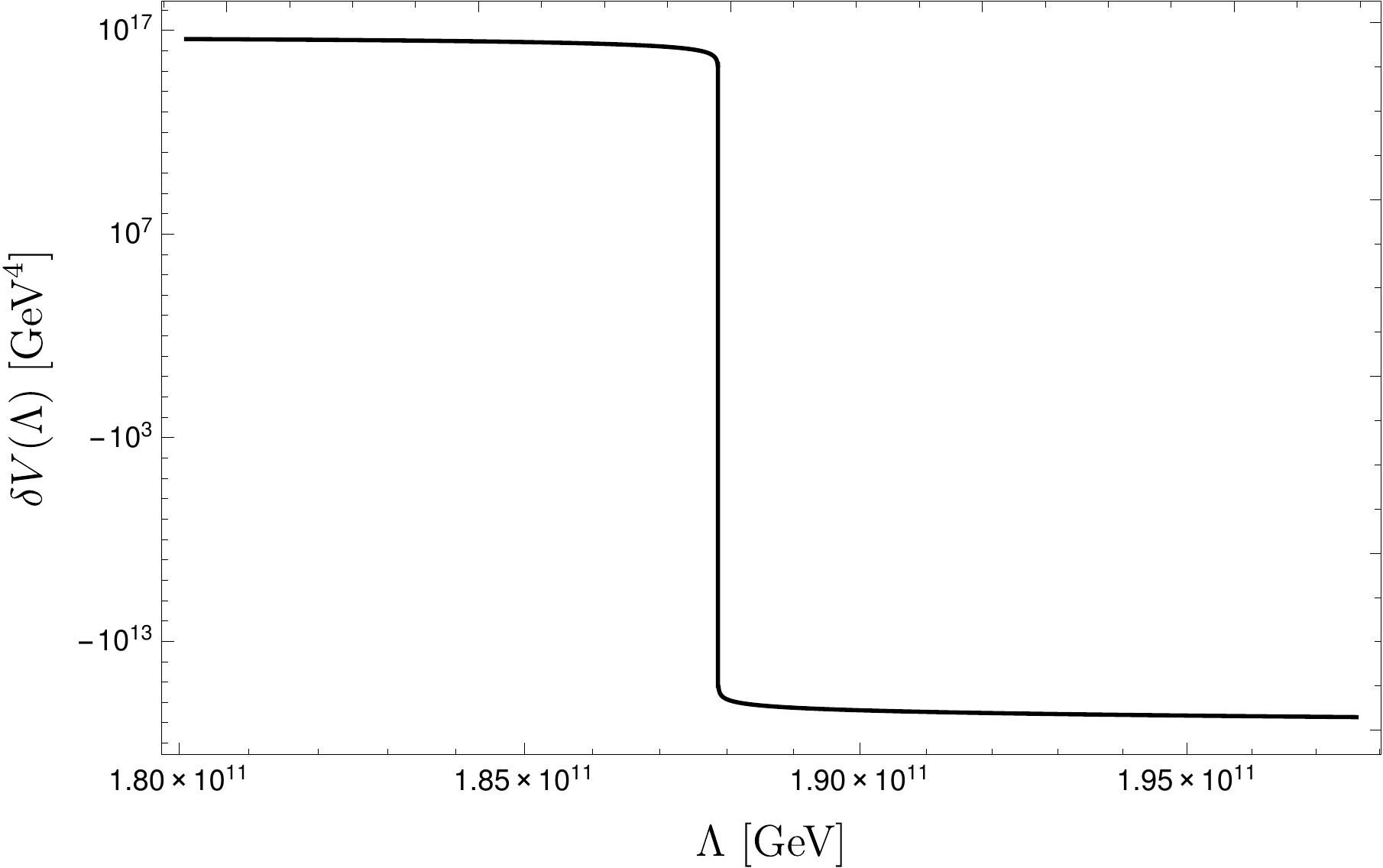}
\caption{The difference $\delta V :=V^\Lambda_{\text{SM}} (\minUV) - V^\Lambda_{\text{SM}} (\minEW)$ of values of the RG improved potential $V^\Lambda_{\text{SM}}$ at two minima as a~function of the scale of new physics $\Lambda$.\label{degeneracy_plot}}
\end{figure}

The difference $\delta V:=V^\Lambda_{\text{SM}} (\minUV) - V^\Lambda_{\text{SM}} (\minEW)$ between values of the RG improved potential $V^\Lambda_{\text{SM}}$ at two minima as a~function of the suppression scale $\Lambda$ is presented in the figure \ref{degeneracy_plot}. The fine-tuning problem is made evident in this plot by the steep character of the presented function around the scale $\Lambda_{\text{deg}}=1.88 \times 10^{11}\; \textrm{GeV}$. 

Fine-tuning of the value of $\lambda_6$ coupling constant poses severe problems for numerical methods. Firstly, in order to calculate the value of the RG improved potential with fine-tuned numerical value of running coupling constants one is required to numerically solve RGEs of the SM with very high precision. Furthermore, the evolution of long-lived networks of domain walls in the case of nearly degenerate minima is difficult to simulate numerically, because it requires a large number of integration steps leading to accumulation of truncation errors.

\section{Properties of SM domain walls in the presence of the $h^6$ operator\label{wall}}
We are interested in the evolution of the time and space dependent expectation value $\langle h(x) \rangle$ of the Higgs field $h$. By definition, an expectation value of a field strength $\Psi_{cl}(x)$ satisfies the equation of motion: $\frac{\delta}{\delta \Psi_{cl}(x)} \Gamma_{\text{eff}}[\Psi_{cl}]=0$, where $\Gamma_{\text{eff}}$ is the 1PI effective action. The 1PI effective action $\Gamma_{\text{SM}}$ of the SM is not known exactly and we approximated it by the following expression:
\begin{equation}
\Gamma_{\text{SM}}[\phi] \approx \int d^4x \sqrt{|\det g|} \left[ \frac{1}{2} g^{\mu \nu} \partial_\mu \phi \partial_\nu \phi -V^\Lambda_{\text{SM}}(\phi) \right], \label{action_spproximation}
\end{equation}
where $\phi$ is a~real scalar field which models the Higgs field in our simulations and $V^\Lambda_{\text{SM}}$ is the effective potential of the SM in zero temperature with the nonrenormalizable operator $h^6$ included. 
Assuming the gravitational background in the form of the Friedman-Robertson-Walker metric background:
\begin{equation}
g = dt^2 - a^2(t) \delta_{ij} dx^i dx^j = a^2(\eta) \left(d\eta^2 -\delta_{ij} dx^i dx^j\right), 
\end{equation}
where Latin indices correspond to spatial coordinates, $t$ is cosmic time and $\eta$ denotes conformal time (such that $d \eta = \frac{1}{a(t)} dt$), the eom for the approximated effective action from eq. \eqref{action_spproximation} can be written down in the form:
\begin{equation}
\frac{\partial^2 \phi}{\partial \eta^2} + \frac{2}{a} \left(\frac{d a}{d \eta}\right) \frac{\partial \phi}{\partial \eta} - \Delta \phi + a^2 \frac{\partial V^\Lambda_{\text{SM}}}{\partial \phi}=0. \label{SM_eom}
\end{equation}

We have solved eq. \eqref{SM_eom} in numerical simulations on the 3D lattice, based on the PRS algorithm \cite{Press:1989yh}. To properly model dynamics of a network of domain walls in these simulations, one needs to know the width of a typical domain wall $w^\Lambda$. In the past, simulations with the physical width of walls varying from 2 to 100 lattice spacing $l$ (i.e. the physical distance between neighbouring points) were performed \cite{Press:1989yh,Coulson:1995uq,Lalak:1996db,Oliveira:2004he,Lalak:2007rs,Kawasaki:2011vv,Leite:2011sc,Hiramatsu:2013qaa}.

In our previous article \cite{Krajewski:2016vbr} we have proposed the algorithm for computing the width of domain walls for a~generic potential. We started with the observation that the time independent, planar (i.e. translationally invariant in two directions) solution $\varphi(x)$ of eq. \eqref{SM_eom} in Minkowski background ($a(t)=const=1$) is given in the implicit, integral form: 
\begin{equation}
x^\Lambda (\varphi_2)-x^\Lambda (\varphi_1)= \int_{\varphi_1}^{\varphi_2} \frac{d\varphi}{\sqrt{2\left(V^\Lambda_{\textrm{SM}}\left(\varphi\right)-V^\Lambda_{\textrm{SM}}(\minEW)\right)}}.\label{approx_solution}
\end{equation}
Furthermore the potential energy density $\sigma(x_1,x_2)$ of the solution \eqref{approx_solution} is given by the following integral: 
\begin{equation}
\sigma^\Lambda(x_1,x_2):=\int_{x_1}^{x_2} V^\Lambda_{\textrm{SM}}(\varphi(x)) dx = \int_{\varphi(x_1)}^{\varphi(x_2)}\frac{V^\Lambda_{\textrm{SM}}(\varphi) d\varphi}{\sqrt{2\left(V^\Lambda_{\textrm{SM}}\left(\varphi\right)-V^\Lambda_{\textrm{SM}}(\minEW)\right)}}. \label{tension}
\end{equation}

\begin{figure}[t]
\centering
\includegraphics[width= 0.7\textwidth]{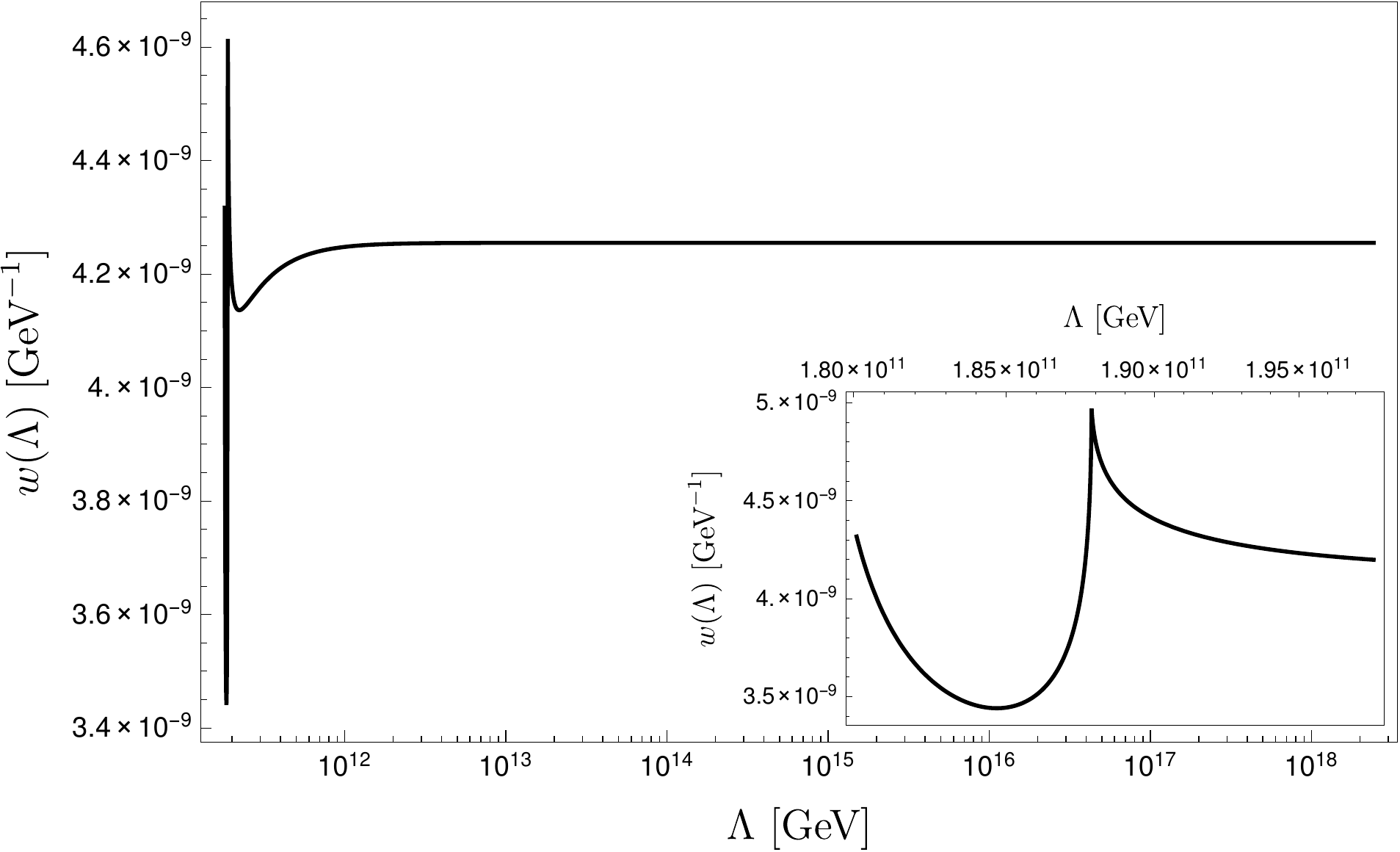}
\caption{The width of domain walls $w$ as a~function of the scale of new physics $\Lambda$.\label{width_plot}}
\end{figure}

Finally, we have found values $\tilde{\varphi_1}$ and $\tilde{\varphi_2}$ such that $V^\Lambda_{\textrm{SM}}\left(\tilde{\varphi_1}\right)=V^\Lambda_{\textrm{SM}}\left(\tilde{\varphi_2}\right)$ and the majority of the potential energy density of the solution \eqref{approx_solution} is stored between $x(\tilde{\varphi_1})$ and $x(\tilde{\varphi_2})$:
\begin{equation}
\frac{\sigma^\Lambda(x(\tilde{\varphi_1}),x(\tilde{\varphi_2}))}{\sigma^\Lambda(-\infty,+\infty)}\approx 97 \%.
\end{equation}
Our estimation of the width of domain walls is then given by:\footnote{We are using the convention $\hbar=1=c$, so lengths and times are expressed in units of $\textrm{GeV}^{-1}$.}
\begin{equation}
w (\Lambda) :=x^\Lambda (\tilde{\varphi_2})-x^\Lambda (\tilde{\varphi_1}). \label{SM_width}
\end{equation}
The estimated width of domain walls as a~function of the suppression scale $\Lambda$ of the nonrenormalizable operator $h^6$ is presented in the figure \ref{width_plot}. Resulting values of the width lay in the range $3.5 \times 10^{-9}\; \textrm{GeV}^{-1} \le w(\Lambda) \le 5.0 \times 10^{-9}\; \textrm{GeV}^{-1}$.

We have chosen the physical lattice spacing $l$ to be equal to $10^{-10}\; \textrm{GeV}^{-1}$ which leads to widths of walls contained in the range $35\, l \le w(\Lambda) \le 50\, l$. Henceforth, we will often use $\frac{1}{l} = 10^{10}\; \textrm{GeV}$ as a~unit of energy and inverse distance in space-time.

\section{Initial conditions for simulations\label{initial}}
Inflationary models predict that the Higgs field expectation value during inflation can be statistically described by the probability distribution $P(h)$ whose evolution satisfies the Fokker-Planck equation. Solutions of the Fokker-Planck for the Higgs field \cite{Coulson:1995uq,Espinosa:2007qp,Hook:2014uia,Espinosa:2015qea,East:2016anr}, in the first approximation, are of the form of the Gauss distribution:
\begin{equation}
P(h) = \frac{1}{\sqrt{2 \pi} \sigma_I} e^{-\frac{\left(h -\theta\right)^2}{2 \sigma_I^2}}, \label{gauss_distribution}
\end{equation}
with the standard deviation of the order of
\begin{equation}
\sigma_I \sim \frac{\sqrt{N} H_I}{2 \pi},
\end{equation}
where $H_I$ is the value of the Hubble constant during inflation and $N$ is the number of e-folds. The Hubble constant $H_I$ is the main parameter of inflationary models. Solving majority of problems found in the standard cosmology requires at least from $50$ to $60$ e-folds, so the standard deviation $\sigma_I$ can be determined in an~inflationary model. Unfortunately, evolution of $P(h)$ turns out to be independent of the mean value of the Higgs field and this value had not changed significantly during inflation. As a result the mean value of the Higgs field after inflation $\theta$ cannot be determined in an~inflationary model and a~priori it can take any value from $0$ to the Planck scale.

We focus on the evolution of networks initialised with $\theta=0$. This choice leads to the least conservative bounds on inflationary models. The lower the value of $\theta$ is, the higher the fraction of lattice points in $\EWbasin$ is and consequently the more likely it is that the evolution of the network will end up in the electroweak vacuum.

Our simulations were initialised at the conformal time $\eta_{start}$ ranging from \mbox{$10^{-2}\, l = 10^{-12}\; \textrm{GeV}^{-1}$} to \mbox{$1 \, l = 10^{-10}\; \textrm{GeV}^{-1}$}. In our previous work~\cite{Krajewski:2016vbr} we showed that taking initial times below this interval does not modify the results. On the other hand, cosmological domain walls whose evolution we model in our numerical simulations, need to be superhorizon at the initialisation (i.e. width of domain walls must be larger than the Hubble horizon $H(\eta_{start})^{-1} \sim a(\eta_{start}) \eta_{start}$ at the time of initialisation $\eta_{start}$). The conformal time when domain walls are formed in the early Universe, must be greater than $\eta_{start}$ for the the initial numerical fluctuations to be smoothed out by the evolution of the field.

\section{Decay of domain walls\label{decay_time}}
In our previous work \cite{Krajewski:2016vbr} we investigated the evolution of networks of Higgs domain walls in the SM valid up to the Planck scale. We showed that obtaining a network that collapses to the electroweak vacuum is not as difficult as one may expect, although it requires certain tuning of initial conditions. Our simulations predict that only a slight dominance of the electroweak vacuum in the initial configuration is required in order for the evolution to end in this vacuum. Previously we concluded that only initial probability distributions \eqref{gauss_distribution} with $\sigma$ lower then $3.25 \times 10^{10}\; \textrm{GeV}$ ($\theta=0$) produce networks that decay to the electroweak vacuum.
Unfortunately, this value of $\sigma$ is not a~gauge-independent quantity, however, the fraction of space occupied by the Higgs field strength belonging to $\EWbasin$ is a~gauge-independent quantity \cite{Espinosa:2015qea}. Moreover, we found that networks of Higgs domain walls are highly unstable. The estimated time of their decay ranges from \mbox{$8 \times 10^{-11}\; \textrm{GeV}^{-1} = 0.8\,l$} to \mbox{$3 \times 10^{-9}\; \textrm{GeV}^{-1} = 30\, l$}. Such short lifetimes exclude a~scenario in which SM domain walls dominate the Universe leading to large distortion of the Cosmic Microwave Background Radiation.

\begin{figure}[!t]
\subfloat[]{\label{time_-2_volume}
\includegraphics[width=0.5 \textwidth]{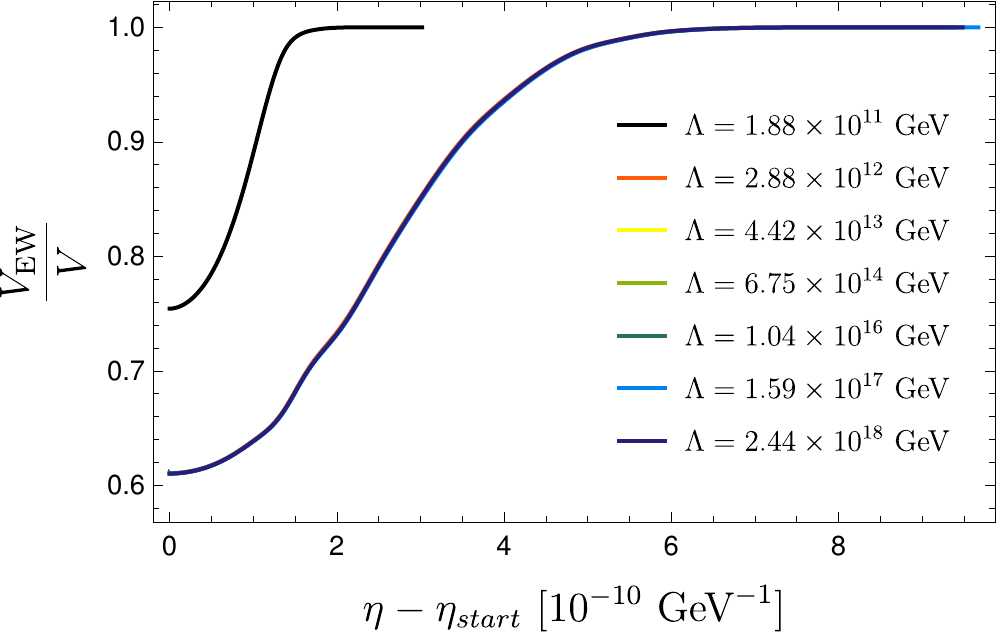}
 }
 \subfloat[]{\label{time_+0_volume}
\includegraphics[width=0.5 \textwidth]{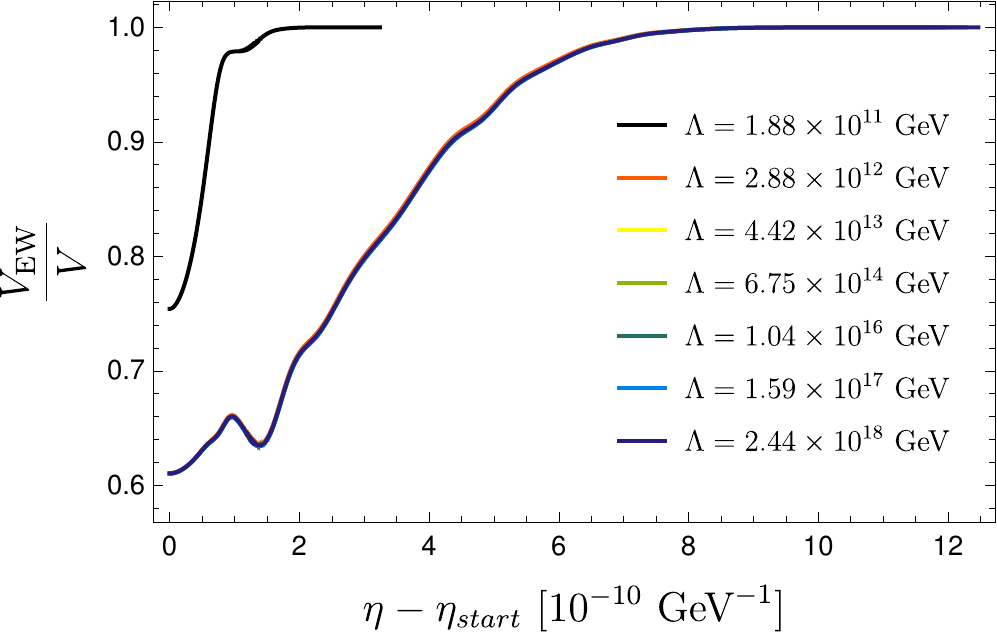}
 }
\caption{Time dependence of the fraction of lattice sites occupied by the field on the electroweak side of the barrier $\frac{V_{\text{EW}}}{V}$ for two different values of the conformal initialisation time \mbox{$\eta_{start}=10^{-12}\; \textrm{GeV}^{-1}$ \protect\subref{time_-2_volume}} and \mbox{$\eta_{start}=10^{-10}\; \textrm{GeV}^{-1}$ \protect\subref{time_+0_volume}} and the standard deviation $\sigma=3.25 \times 10^{10}\; \textrm{GeV}$ at the initialisation.\protect\label{evolution}}
\end{figure}

The main aim of this work is to generalize our previous results to include BSM physics. As we discussed in section \ref{SM_potential}, new interactions beyond the SM may significantly change the form of the Higgs effective potential. These modification influence the evolution of networks of Higgs domain walls. In this section we present results of our new numerical simulations. We study dependence of the decay time of networks on the suppression scale $\Lambda$ of the nonrenormalizable operator $h^6$. An~example of the time evolution of such a network is presented in the figure \ref{evolution}. In this figure the ratio $\frac{V_{\text{EW}}}{V}$ of number of lattice sites occupied by field strengths in $\EWbasin$ to the total volume of the lattice, as a~function of conformal time $\eta$, is plotted for two different values of the conformal initialisation time $\eta_{start}$. Results calculated for various values of the suppression scale $\Lambda$ and the same initialisation conditions have been arranged.
 Presented results for each case were averaged over five simulations on the lattice of the size $256^3$.

A~few observations can be deduced from presented plots. Firstly, the evolution of domain walls quickly becomes insensitive to corrections from nonrenormalizable operators for the suppression scale larger than $\Lambda_{deg} = 1.88 \times 10^{11}\,\textrm{GeV}$ for which the minima of the potential are degenerate. This conclusion is supported by results of our scan over the space of the initialisation parameter $\sigma$.

\begin{figure}[!t]
\subfloat[]{\label{time_-2_big}
\includegraphics[width=0.5 \textwidth]{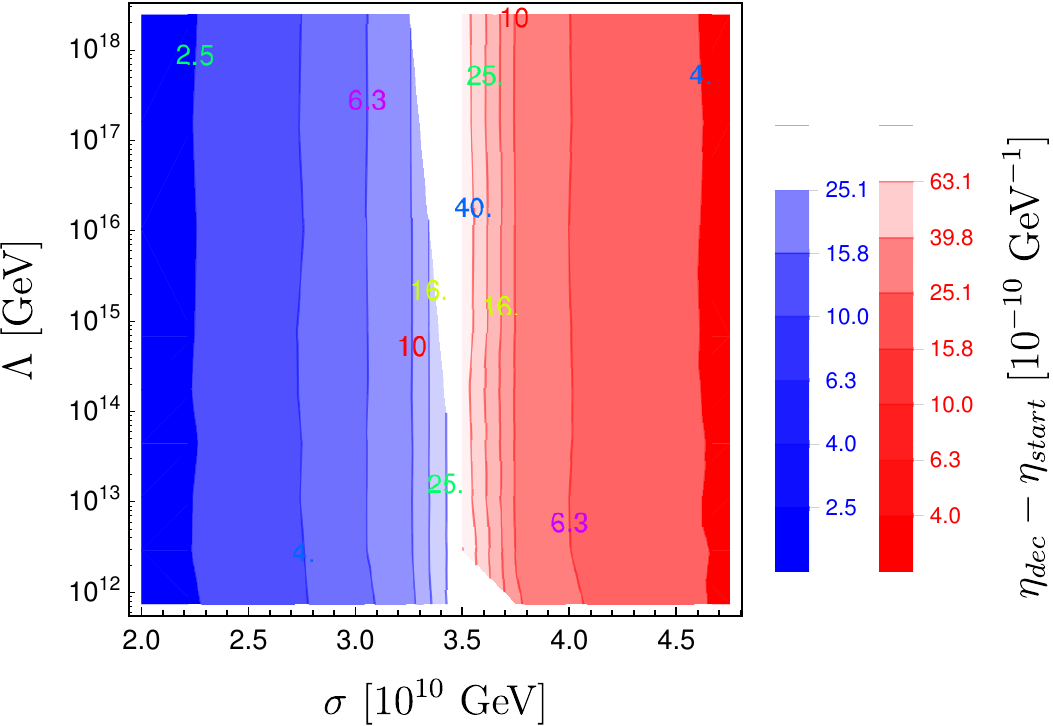}
 }
 \subfloat[]{\label{time_+0_big}
\includegraphics[width=0.5 \textwidth]{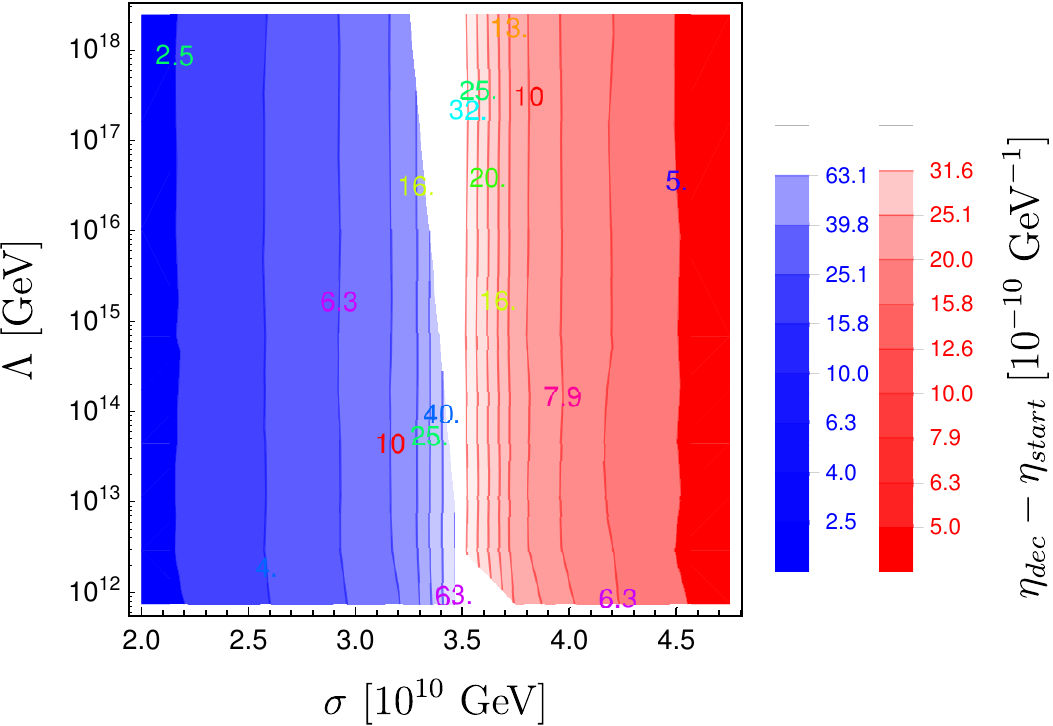}
 }
\caption{Dependence of the decay time $\eta_{dec} - \eta_{start}$ of networks of Higgs domain walls as a~function of the standard deviation $\sigma$ of the initialisation probability distribution and the suppression scale $\Lambda$ of the nonrenormalizable operator $h^6$ for two different values of the conformal initialisation time \mbox{$\eta_{start}=10^{-12}\; \textrm{GeV}^{-1}$ \protect\subref{time_-2_big}} and \mbox{$\eta_{start}=10^{-10}\; \textrm{GeV}^{-1}$ \protect\subref{time_+0_big}}. Blue regions correspond to networks decaying to the EWSB vacuum and red ones to networks decaying to the high field strength minimum.\protect\label{decay_time_big}}
\end{figure}

Estimated dependence of the conformal decay time of the network on the value of the standard deviation $\sigma$ at the initialisation and the value of the suppression scale $\Lambda$ is plotted in the figure \ref{decay_time_big}. Blue regions in these plots were extrapolated from the simulations in which evolution of networks ended in the basin of the attraction of the EWSB vacuum $\EWbasin$ and red ones from networks decaying into the basin of the attraction of the high field strength minimum (excluded by experiments).

Nearly vertical shape of resulting contours proves that dependence on the value of suppression scale $\Lambda$ of the decay time of networks is very weak in the shown range of $\Lambda$. Moreover, a quick inspection of plots in the figure \ref{decay_time_big} reveals that late domain walls, i.e. initialised at $\eta_{start} = 1\, l =10^{-10}\; \textrm{GeV}^{-1}$, have the longer decay time what is consistent with our previous findings \cite{Krajewski:2016vbr}. 

\begin{figure}[!t]
\subfloat[]{\label{time_-2_small}
\includegraphics[width=0.5 \textwidth]{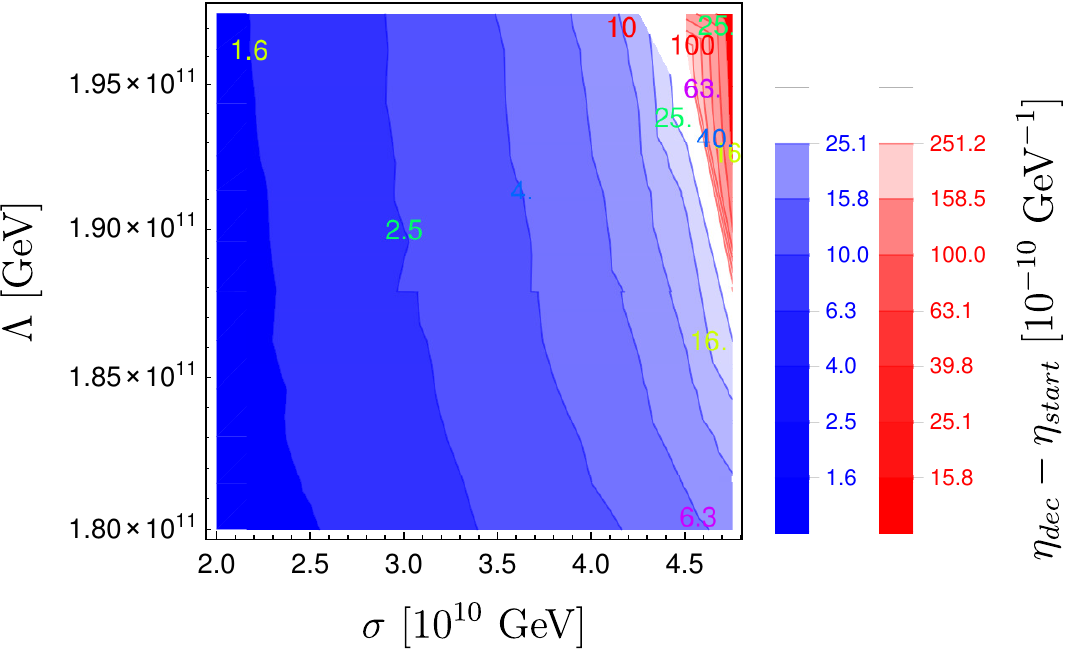}
 }
 \subfloat[]{\label{time_+0_small}
\includegraphics[width=0.5 \textwidth]{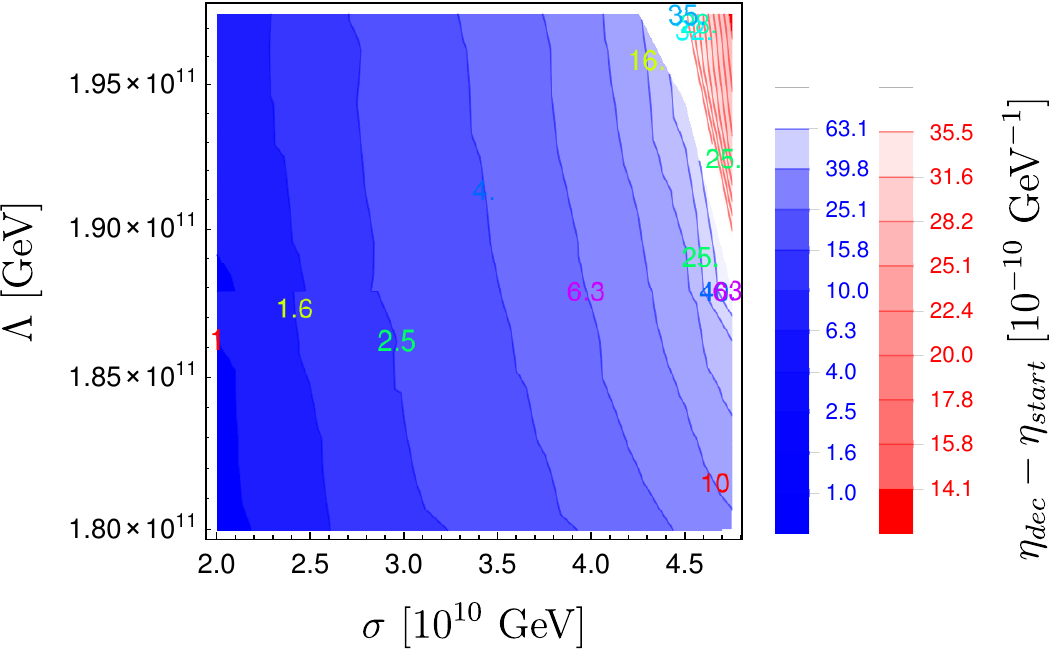}
 }
\caption{Dependence of the decay time $\eta_{dec} - \eta_{start}$ of networks of Higgs domain walls as a~function of the standard deviation $\sigma$ of the initialisation probability distribution and the suppression scale of the nonrenormalizable operator $h^6$ for two different values of the conformal initialisation time \mbox{$\eta_{start}=10^{-12}\; \textrm{GeV}^{-1}$ \protect\subref{time_-2_small}} and \mbox{$\eta_{start}=10^{-10}\; \textrm{GeV}^{-1}$ \protect\subref{time_+0_small}}. Blue regions correspond to networks decaying to the EWSB vacuum and red to networks decaying to the high field strength minimum.\protect\label{decay_time_small}}
\end{figure}

Secondly, from plots in the figure \ref{evolution} we see that the decay of networks driven by the effective potential with nearly degenerate minima proceeds differently than it would with higher suppression scales $\Lambda$. The initial value of the fraction $\frac{V_{\text{EW}}}{V}$ in the case of \mbox{$\Lambda=\Lambda_{deg}=1.88 \times 10^{11}\; \textrm{GeV}$} is higher than for higher values of the suppression scale $\Lambda$. Moreover, networks of domain walls in this case decay in the conformal time at least three times shorter. In the figure \ref{decay_time_small} we present lifetime of networks of domain walls driven by the effective potential with the nonrenormalizable operator suppressed by the scale $\Lambda$ close to the value $\Lambda_{deg}$ at which the minima are degenerate.

\begin{figure}[t]
\centering
\includegraphics[width= 0.7\textwidth]{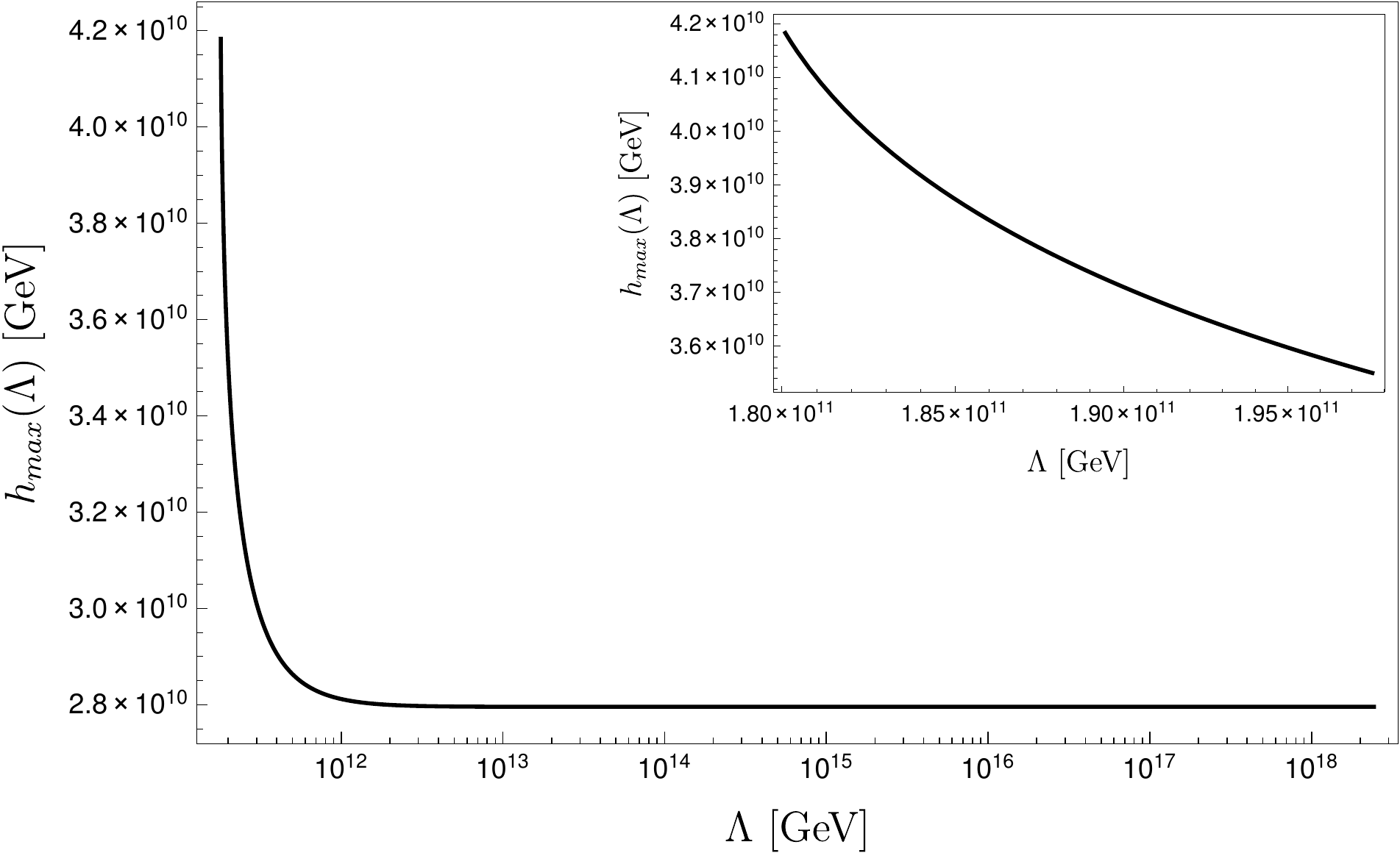}
\caption{The position $h_{max}$ of the local maximum separating two minima of the RG improved effective potential as a~function of the scale of new physics $\Lambda$.\label{maximum_plot}}
\end{figure}

First observation from the figure \ref{decay_time_small} is such that for $\Lambda<\Lambda_{deg}$ the final state of evolution of networks initialised with vanishing mean vale $\theta$ is always the electroweak vacuum. Moreover decay time of networks decreases with the decreasing suppression scale $\Lambda$. This behaviour disagrees with the prediction that domain walls should be metastable in the case of a~potential with degenerate minima. This effect is the result of a~change in the position of the local maximum $h_{max}$ which increases as $\Lambda$ decreases. Figure~\ref{maximum_plot} illustrates how $h_{max}$ changes as a~function of the suppression scale $\Lambda$.

\begin{figure}[!t]
\subfloat[]{\label{time_-2_small_rescaled}
\includegraphics[width=0.5 \textwidth]{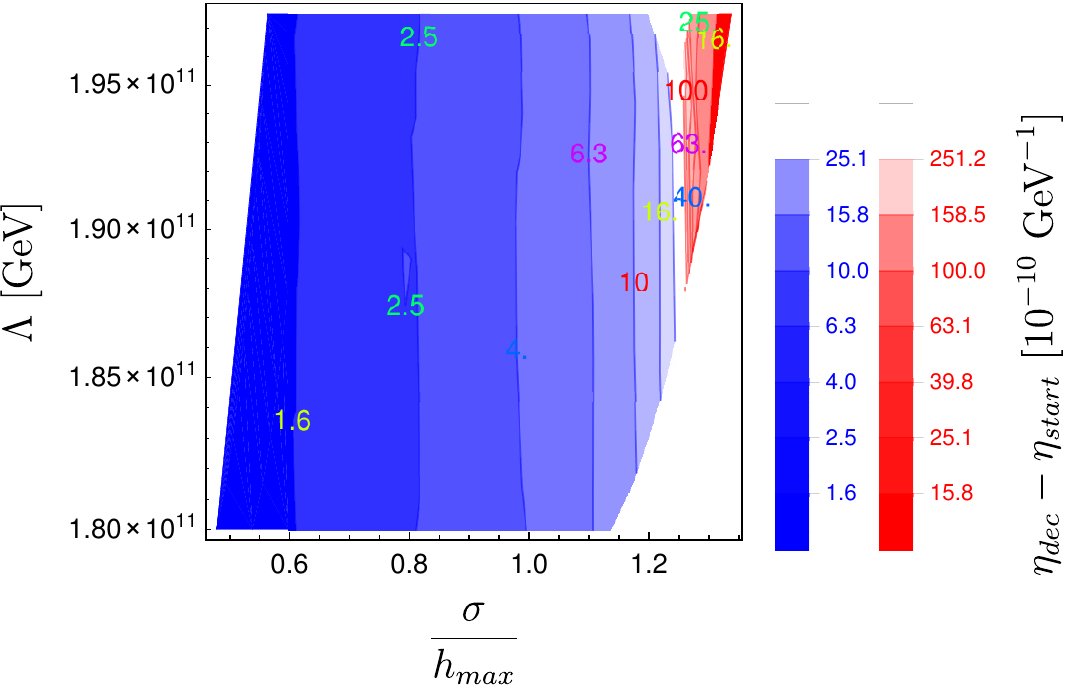}
 }
 \subfloat[]{\label{time_+0_small_rescaled}
\includegraphics[width=0.5 \textwidth]{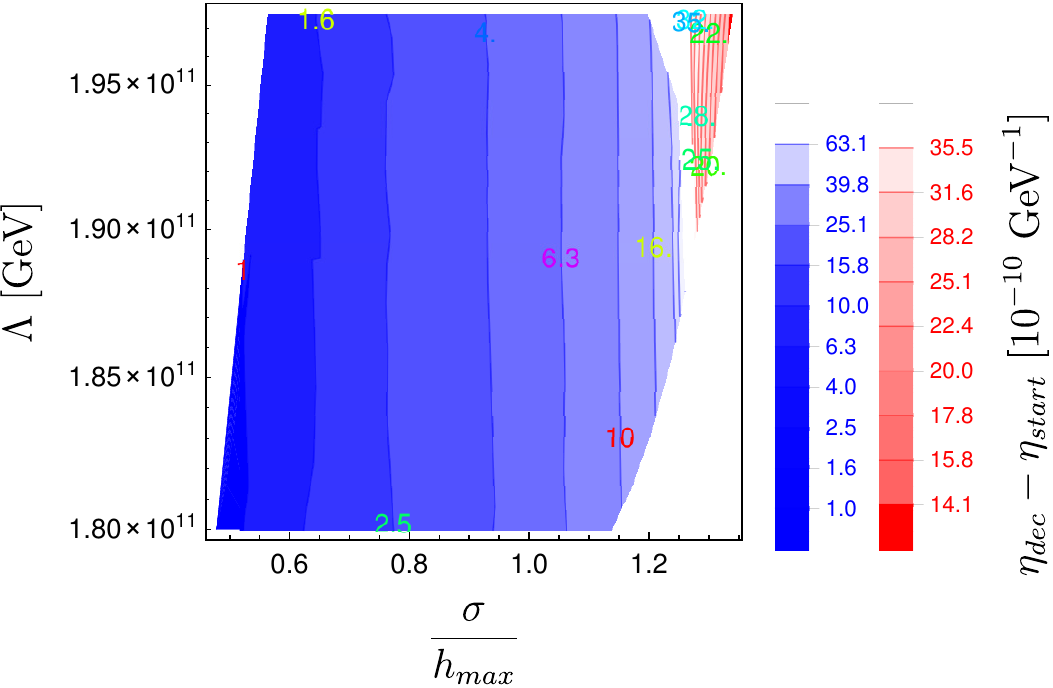}
 }
\caption{Dependence of the decay time $\eta_{dec} - \eta_{start}$ of networks of Higgs domain walls as a~function of the ratio $\frac{\sigma}{h_{max}}$ of the standard deviation of initialisation probability distribution $\sigma$ over the position of the local maximum separating the two minima $h_{max}$ and the suppression scale $\Lambda$ of the nonrenormalizable operator $h^6$ for two different values of the conformal initialisation time \mbox{$\eta_{start}=10^{-12}\; \textrm{GeV}^{-1}$ \protect\subref{time_-2_small_rescaled}} and \mbox{$\eta_{start}=10^{-10}\; \textrm{GeV}^{-1}$ \protect\subref{time_+0_small_rescaled}}. Blue regions correspond to networks decaying to the EWSB vacuum and red to networks decaying to the high field strength minimum.\protect\label{decay_time_small_rescaled}}
\end{figure}

Second result of the shift in $h_{max}$ that can be observed in the figure \ref{evolution} is a change in the starting value of the fraction $\frac{V_{\text{EW}}}{V}$ in the case of $\Lambda=\Lambda_{deg}$. Figure~\ref{decay_time_small_rescaled} shows the decay time of networks of domain walls as a~function of the suppression scale $\Lambda$ and the ratio $\frac{\sigma}{h_{max}(\Lambda)}$. Again, the nearly vertical arrangement of contours indicates weak dependence on the suppression scale $\Lambda$.

\section{Metastable networks decay\label{metastability}}
Our simulations show that for generic initialisation probability distributions ($\theta=0$ and different values of $\sigma$) networks of Higgs domain walls were unstable and decayed shortly after the formation. Moreover, considering two values of the suppression scale $\Lambda$ fine-tuned in order to produce potentials with nearly degenerate minima with difference between values at minima of the order of $\mathcal{O} \left(10^{26} \textrm{GeV}^4\right)$ and $\mathcal{O} \left(10^{10} \textrm{GeV}^4\right)$, we have not observed significant difference in the life-time these networks.

Previous studies predict that networks of cosmological domain walls whose dynamics is driven by potentials with nearly degenerate minima evolved in the so called scaling regime. This means the number of domain walls in the Hubble horizon was preserved during the evolution of the network and the typical size of walls was of the order of the Hubble radius. These networks then became unstable due to a~too large Hubble radius and their evolution ended by decay of the network. This behaviour was confirmed in many numerical lattice simulations \cite{Press:1989yh,Garagounis:2002kt,Oliveira:2004he,Avelino:2005pe,Avelino:2005kn,Lalak:2007rs,Kawasaki:2011vv,Leite:2011sc,Leite:2012vn,Martins:2016ois}.

However, majority of simulations preformed in the past have dealt with spontaneous breaking of approximate global symmetries. In these scenarios the difference of values of potentials between their minima are generated by small corrections to symmetric potentials. Thus nearly degenerate minima implicate nearly symmetric potential in these situations. Obviously this is not the case of the SM with nonrenormalizable operators. The RG improved effective potential of the SM with the nonrenormalizable operator is asymmetric, even when minima are degenerate. We estimate the typical scale of the modulus of the derivative of the potential on both sides of the local maximum separating the minima (outside a~close neighborhood of the maximum) to be of the order of:
\begin{displaymath}
\left| \frac{\partial \widetilde{V}_{\text{SM}}(h)}{\partial h} \right| \approx \lambda_{\text{eff}}(h; h) h^3 + \frac{\partial \lambda_{\text{eff}}(h; h)}{\partial h} h^4 = \left(\lambda_{\text{eff}}(h; h) + \frac{\partial \lambda_{\text{eff}}(h; h)}{\partial \log h}\right) h^3,\\
\end{displaymath}
where we neglected terms proportional to $h^2$ which are subleading for large field strength. The term in the bracket $\left(\lambda_{\text{eff}}(h; h) + \frac{\partial \lambda_{\text{eff}}(h; h)}{\partial \log h}\right)$ can be expressed via the RGEs in terms of running coupling constants of the SM, thus it depends logarithmically on $h$ far enough from its zero (i.e. the position of the maximum of $\widetilde{V}_{\text{SM}}^\Lambda$). The potential derivative behaves as $h^3$, so that it grows with increasing field strength $h$. As a~result the RG improved effective potential is steeper on high field strengths side of the local maximum. This asymmetry of the potential is the source of the indicated discrepancy between our simulations of the evolution of Higgs domains walls and previously performed simulations of domain walls for spontaneously broken approximated global symmetries.

Previous numerical studies of dynamics of cosmological domain walls \cite{Lalak:2007rs} revealed that asymmetry of the potential can be compensated by asymmetry of the initial configuration of the field leading to metastable networks. In order to validate this reasoning we have searched for asymmetric initial configurations producing long-lived networks of Higgs domain walls for different values of the suppression scale $\Lambda$. We have used probability distributions with $\sigma = 10 ^{10}\; \textrm{GeV}$ and $\theta$ satisfying relation $\theta + \frac{\sigma}{n} = h_{max}$, where $h_{max}$ is the position of the local maximum, and $n$ is a free parameter. This parametrization of initialisation parameters guarantees that the fraction of lattice sites belonging to $\EWbasin$ depends only on $n$ and is independent of $h_{max}$ (which depends on $\Lambda$). We performed scans with various values of $n$ and we found a significant increase in the decay time of networks for values higher than $n=300$ i.e. highly fine-tuned initial configurations.

\begin{figure}[t]
\centering
\includegraphics[width= 0.7\textwidth]{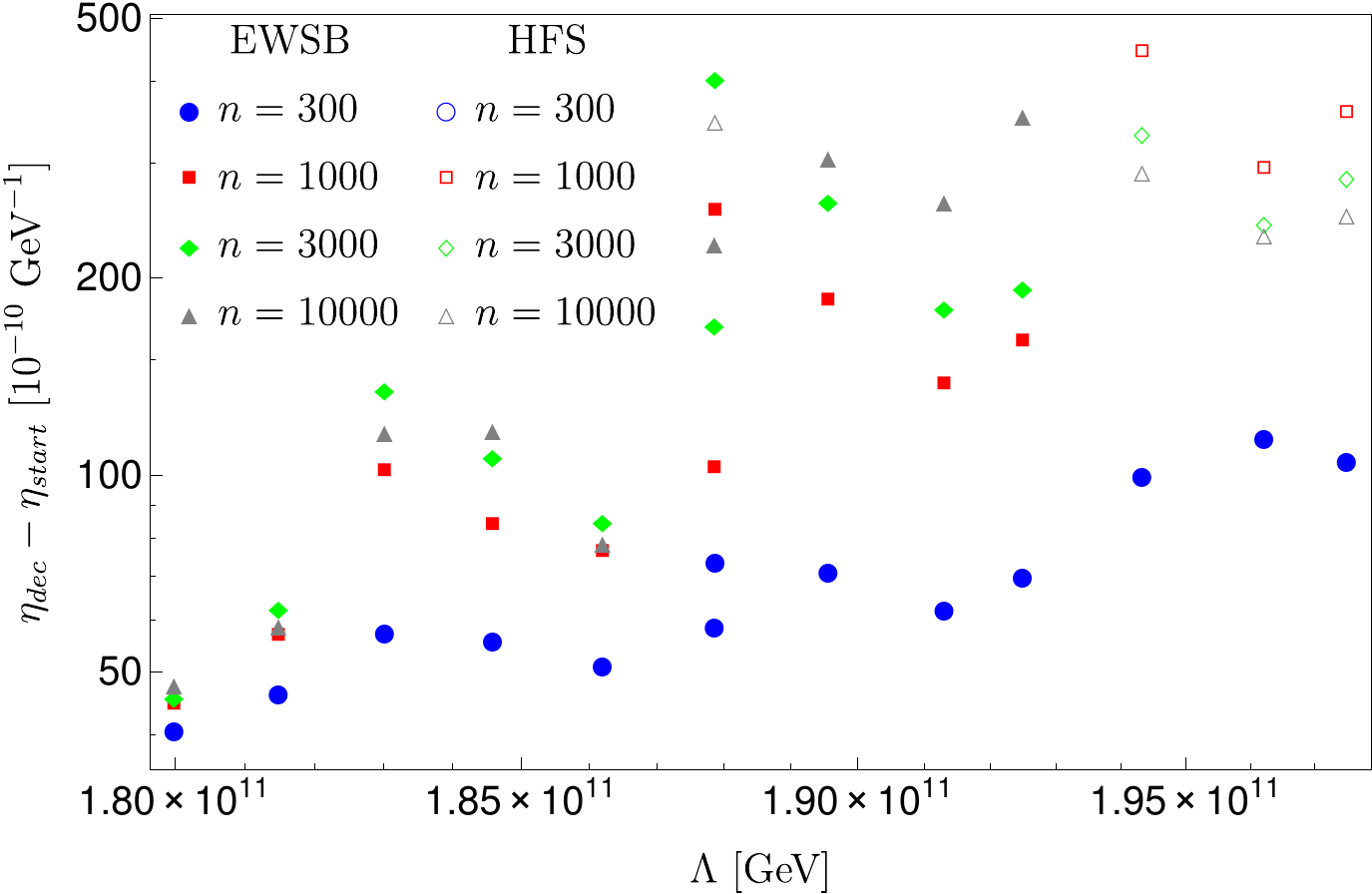}
\caption{Decay times of networks initialised with probability distributions satisfying \mbox{$\theta + \frac{\sigma}{n} = h_{max}$} with various values of $n$ and $\Lambda$. These specific initial conditions were chosen as an~attempt to obtain metastable networks. Blue circles correspond to $n=300$, red squares to $n=1000$, green triangles to $n=3000$ and gray diamonds to $n=10000$. Filled markers indicate networks that ended their evolution in the EWSB vacuum and empty ones those which ended in the high field strength (HFS) minimum $\minUV$.\label{metastable}}
\end{figure}

Decay times obtained in our simulations initialised with $n=300,1000,3000,10000$ are presented in the figure~\ref{metastable}. Points with filled markers correspond to networks whose evolution ended in the electroweak vacuum and points with empty markers to ones for which the final state was the high field strength minimum. Points marked with blue circles corresponds to $n=300$, with red squares to $n=1000$, green triangles are for $n=3000$ and gray diamonds for $n=10000$. Each point in the figure \ref{metastable} is an~average over five independent simulations.

We observed large fluctuations of values of the decay time $\eta_{dec}$ for long-lived networks. This behaviour can be easily understood. The initial probability distribution is reproduced by the initial configuration on a~finite lattice with finite accuracy. Small fluctuations of the initial configuration introduced by finite accuracy of the initialisation algorithm result in large relative fluctuations of the decay time $\eta_{dec}$ in these fine-tuned scenarios. 

Observed decay times were as long as $5 \times 10^{-8}\; \textrm{GeV}^{-1}=500\, l$ and strongly point towards possibility of producing metastable networks for highly fine-tuned initial configurations. However, obtained long decay times need to be treated with caution, because the late evolution of networks in these simulations extends beyond the time range for which finite lattice simulations are reliable. For conformal times larger than the size of the lattice which for presented results was equal to $256^3$, the modelled patch of the space corresponds to only a~fraction of the Hubble horizon at the end of simulations. Thus, boundaries of the lattice are no longer causally disconnected and their influence on the late evolution of networks may be significant. Furthermore, even if boundary conditions would not disturb significantly the simulated evolution, results of long time simulations can still be inaccurate. In the scaling regime Hubble horizon contains only a~few domain walls and it could happen that the simulated patch accommodates none of them. As a~result the computed decay times may underestimate real life-times of long-lived networks of Higgs domain walls. 

For the value of $n$ equal to $n=300$ all considered networks decayed into the EWSB vacuum. For higher values of $n$, i.e. lower fraction of field values belonging to $\EWbasin$ at the lattice at the initialisation and more fine-tuned initial configurations, networks for certain higher values of suppression scales decayed to the (global) high field strength minimum. For majority of the considered values of the suppression scale $\Lambda$ the higher was the value of $n$ (the more fine-tuned was the initial configuration) the more stable were formed networks. Moreover, networks whose dynamics was driven by the effective potential which global minimum is EWSB minimum were less stable than networks for which the high field strength minimum is the global minimum of the effective potential.

\begin{figure}[t]
\subfloat[]{\label{vol_degenerate}
\includegraphics[width=0.5 \textwidth]{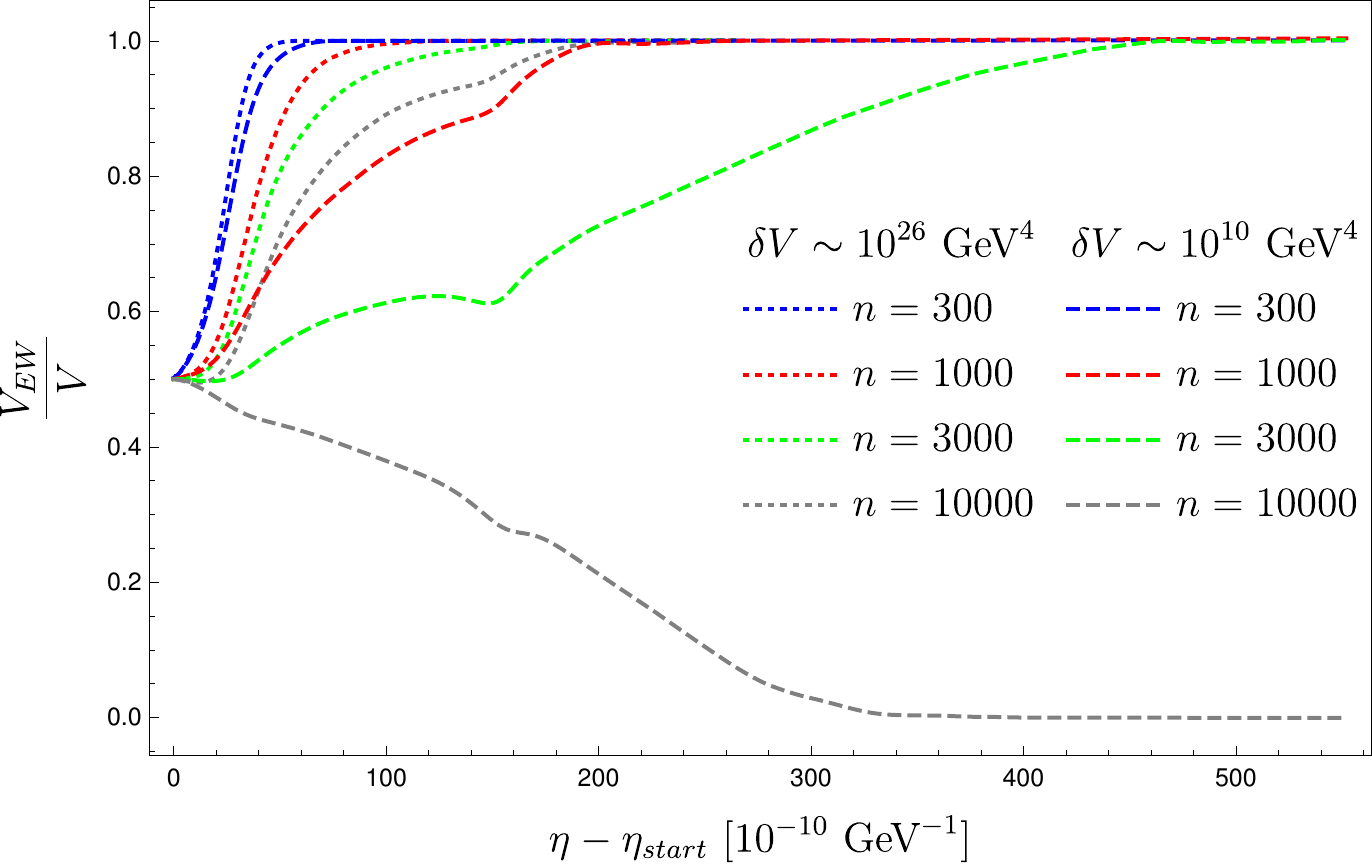}
 }
\subfloat[]{\label{surf_scaling}
\includegraphics[width=0.5 \textwidth]{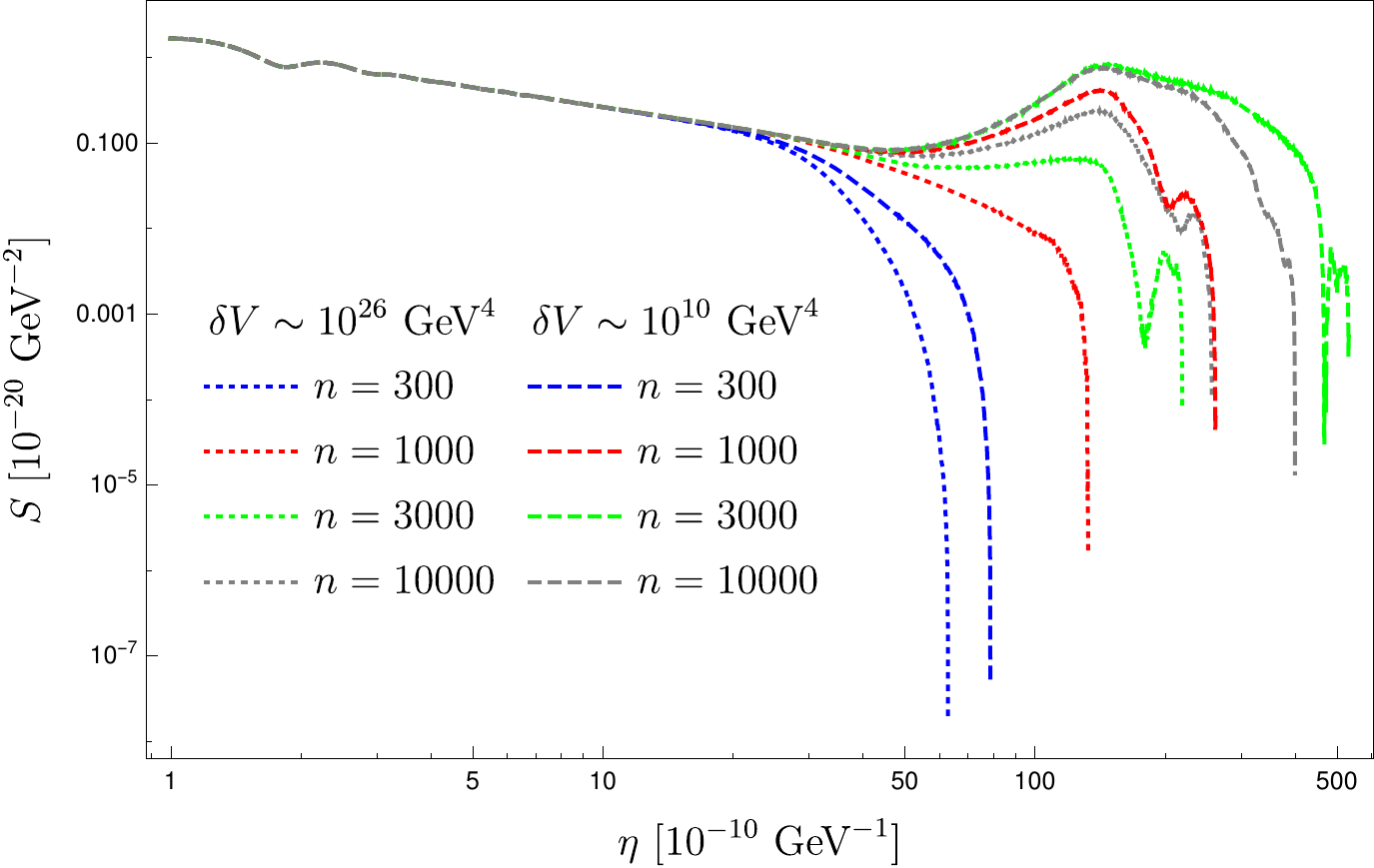}
 }
\caption{Time dependence of the fraction of lattice sites occupied by the field on the electroweak side of the barrier \mbox{$\frac{V_{\text{EW}}}{V}$} \protect\subref{vol_degenerate} and of the surface area of domain walls per lattice site \mbox{\protect\subref{surf_scaling}}. We consider two different precision levels of the fine-tuning of the degeneracy of the minima \mbox{$\delta V = \mathcal{O} \left(10^{26}\; \textrm{GeV}^4\right)$} (dotted) and \mbox{$\delta V = \mathcal{O} \left(10^{10}\; \textrm{GeV}^4\right)$ }(dashed) and four different initialisation probability distributions given by the relation \mbox{$\theta + \frac{\sigma}{n} = h_{max}$} with \mbox{$n=300$} (blue), \mbox{$n=1000$} (red), \mbox{$n=3000$} (green) and \mbox{$n=10000$} (gray). \protect\label{nearly_degenerate}}
\end{figure}

However, for $n=1000$ and $n=3000$ the most stable networks were ones with the effective potential with nearly degenerate minima and the highly fine-tuned difference of the values at the minima. Moreover they were the most stable configurations observed in all our numerical simulations. The time evolution of these networks is presented and compared with the less fine-tuned potential in the figure \ref{nearly_degenerate}. Dotted lines in the figure \ref{nearly_degenerate} correspond to simulations with the effective potential with the difference between values in minima $\delta V = \mathcal{O} \left(10^{26}\; \textrm{GeV}^4\right)$ and dashed lines to ones with values of the potential fined-tuned to $\delta V =\mathcal{O} \left(10^{10}\; \textrm{GeV}^4\right)$. Colours of lines indicate the value of $n$ in the probability distribution used to initialize simulations: blue for $n=300$, red for $n=1000$, green for $n=3000$ and gray for $n=10000$. Plot~\ref{vol_degenerate} shows time dependence of the ratio $\frac{V_{\text{EW}}}{V}$ of lattice sites occupied by the field on the electroweak side of the barrier to the volume of the lattice. Networks of domain walls with the potential with $\delta V = \mathcal{O} \left(10^{26}\; \textrm{GeV}^4\right)$ initialised with $n=300$ and $n=1000$ decayed shortly after the formation. Networks initialised with higher values of $n$, equal to $n=3000$ and $n=10000$, decayed later with life-time increasing for larger $n$. Life-times of networks of domain walls with the potential with $\delta V=\mathcal{O} \left(10^{10}\; \textrm{GeV}^4\right)$ are longer than the ones with the less fined-tuned potential initialised with the same values of $n$. Networks initialised with $n=300$ were not very sensitive to the level of the fine-tuning of $\Lambda$. However, evolution of networks initialised with parameters satisfying relation $\theta + \frac{\sigma}{n} = h_{max}$ with $n \ge 1000$ differ strongly for two considered levels of fine-tuning. Networks initialised with $n=1000$ driven by the potential with $\delta V = \mathcal{O} \left(10^{10}\; \textrm{GeV}^4\right)$ were the most stable networks observed in our numerical simulations.

Better insight into the evolution of long-lived networks is provided by the plot \ref{surf_scaling} which presents time dependence of the surface area of networks. From this plot one may deduce that the evolution of three fastest decaying groups of networks differs qualitatively from the evolution of the rest. The surface area of these three networks decreases monotically with the conformal time $\eta$. On the other hand, in time dependence of surface area of the rest of simulated networks one finds a local maximum followed by the period of slow decay. Such behaviour was previously observed in lattice simulations \cite{Lalak:2007rs} and was connected with networks entering the scaling regime. This behaviour of surface areas is the evidence for Higgs domain walls also evolving in the scaling regime.

\begin{figure}[t]
\subfloat[]{\label{eta=141,29}
\includegraphics[width=0.49\textwidth]{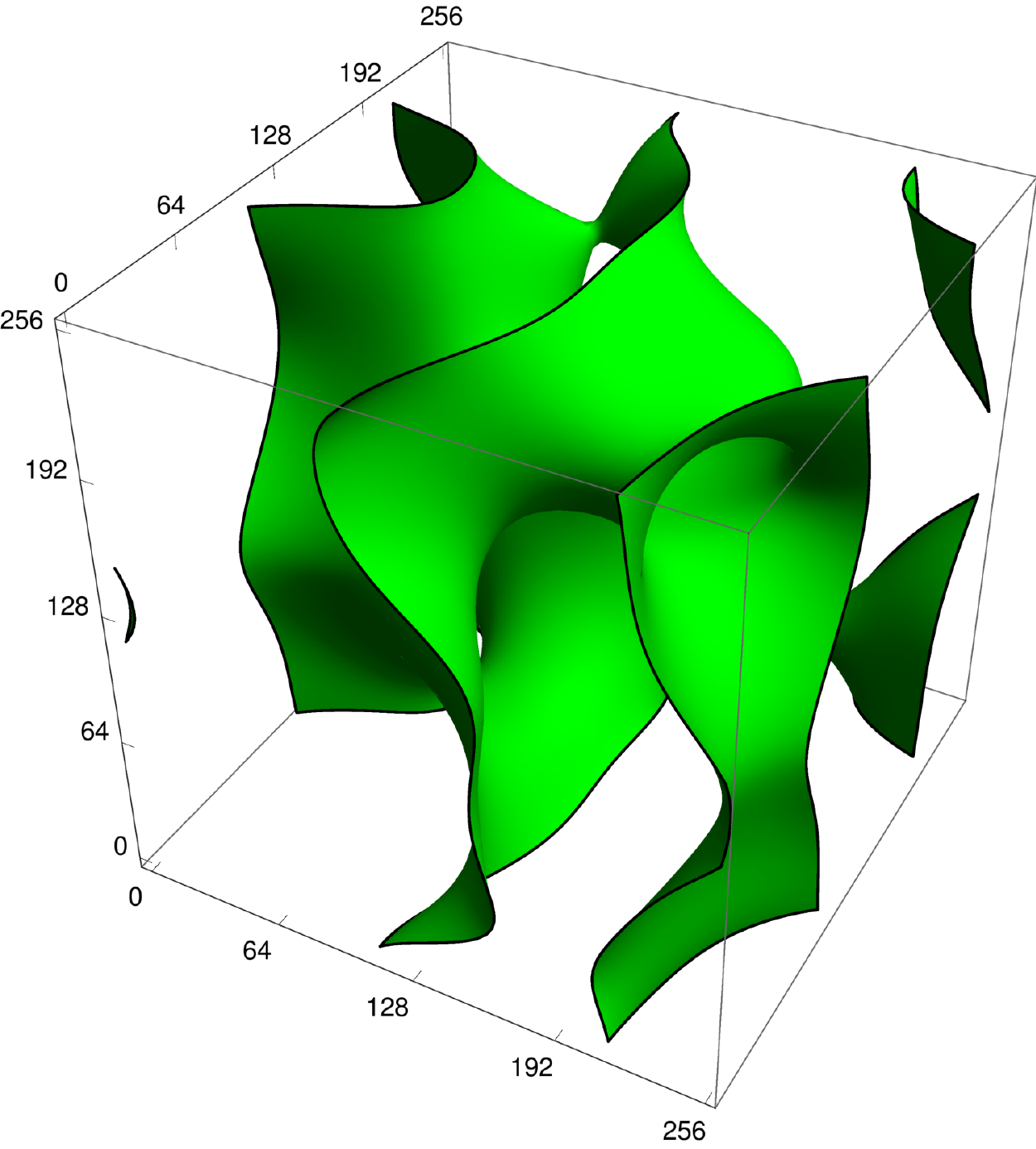}
}
\subfloat[]{\label{eta=211,43}
\includegraphics[width=0.49\textwidth]{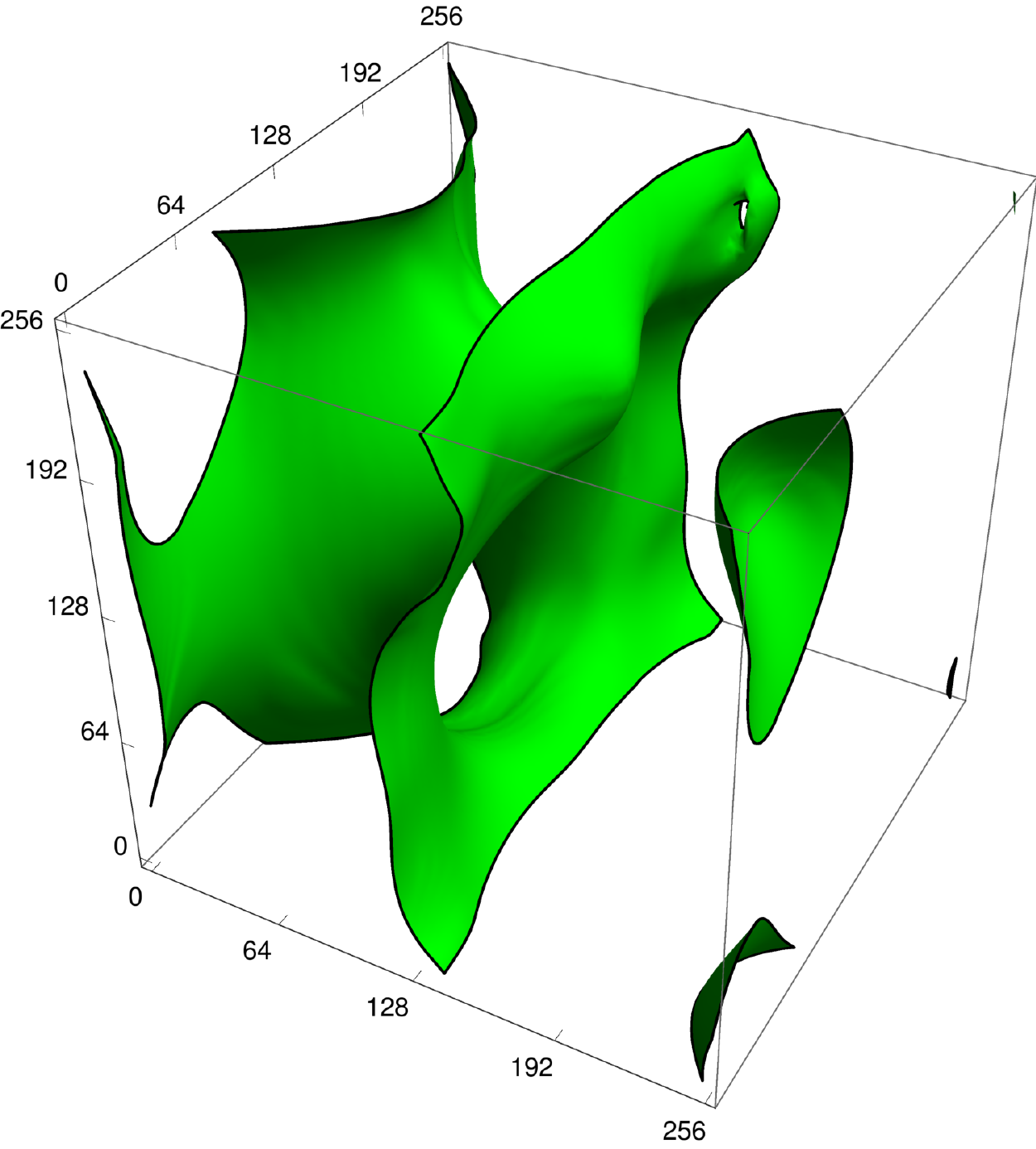}
}\\
\subfloat[]{\label{eta=251,49}
\includegraphics[width=0.49\textwidth]{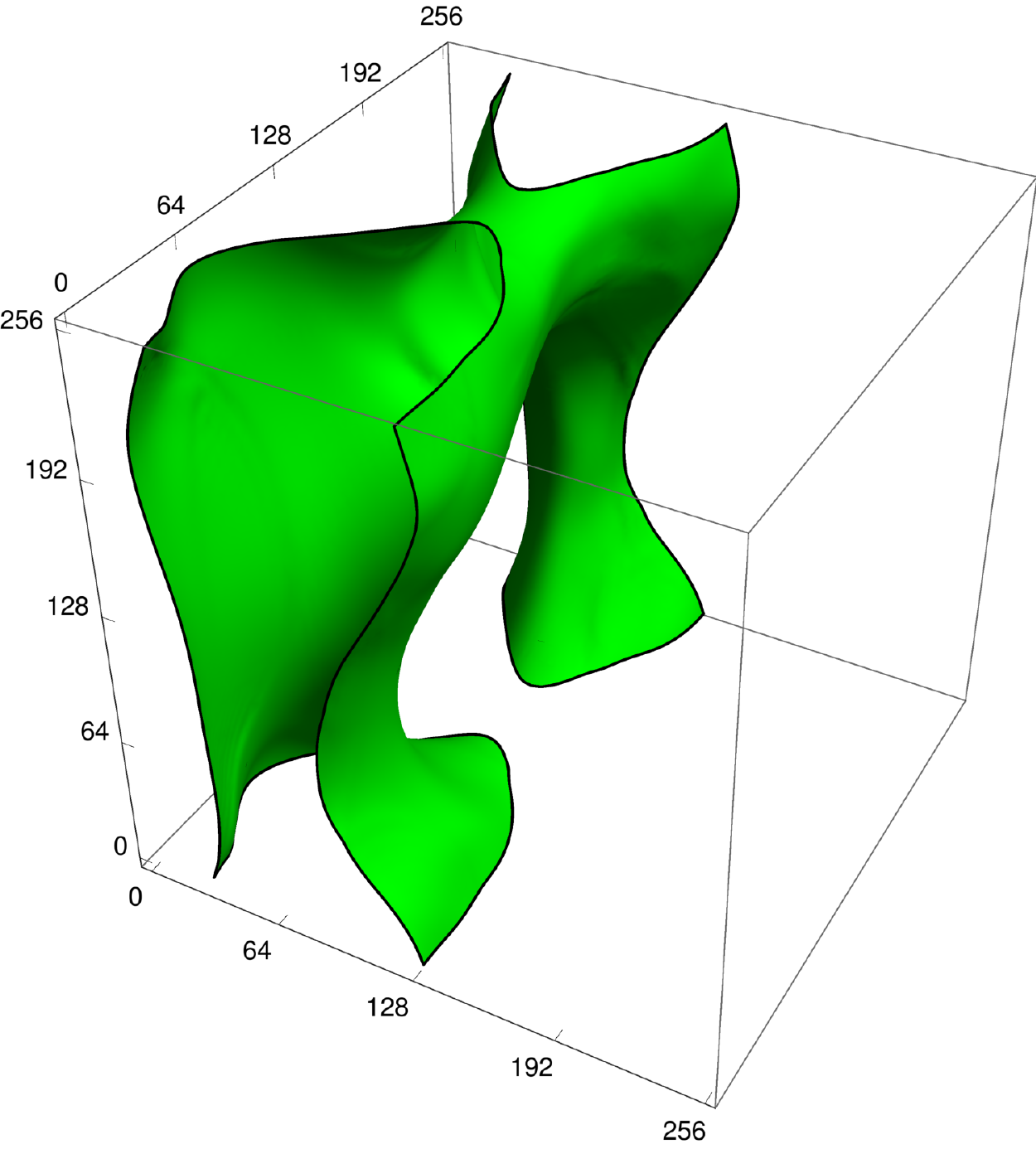}
}
\subfloat[]{\label{eta=301,55}
\includegraphics[width=0.49\textwidth]{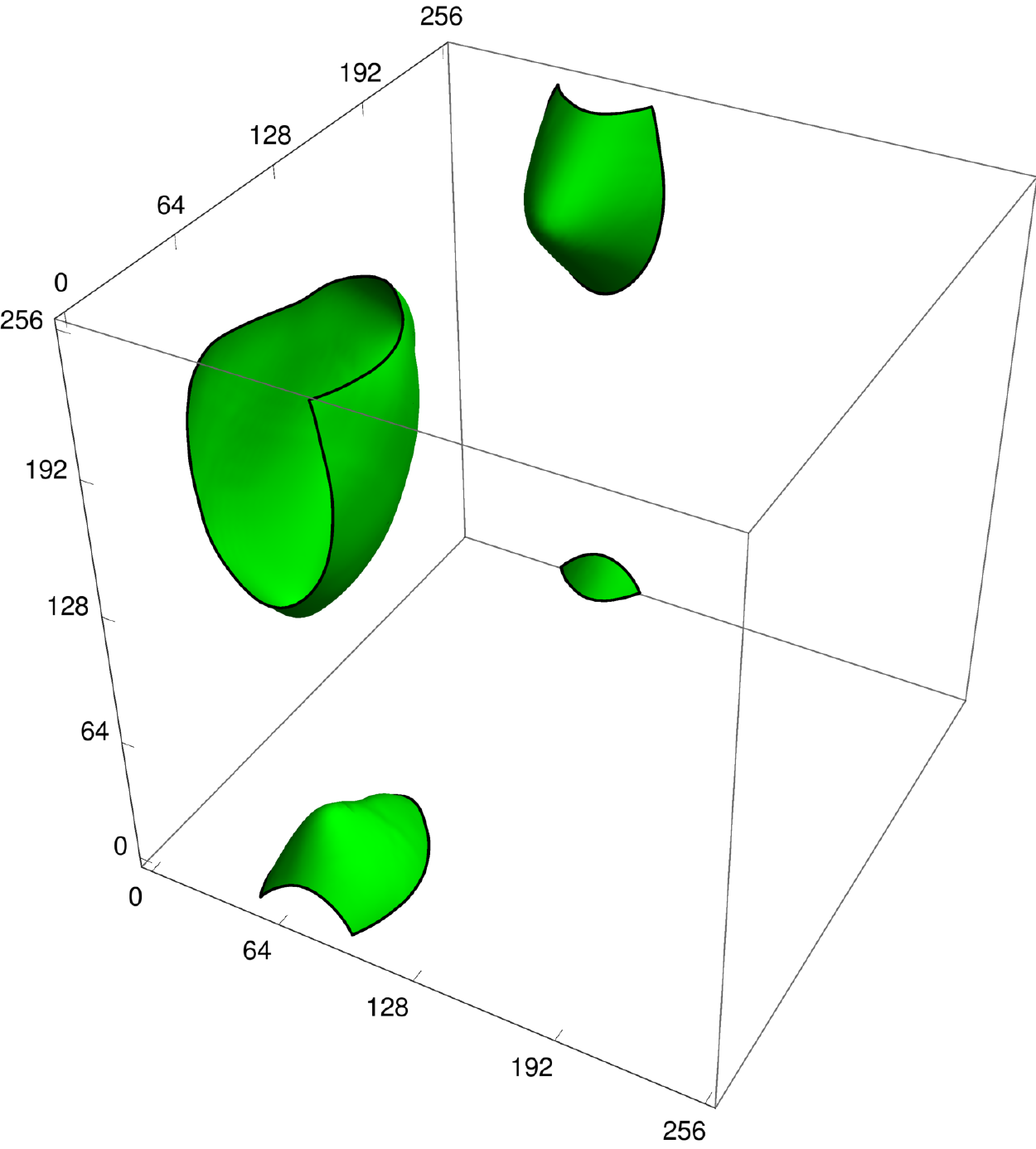}
}
\caption{Visualization of the isosurface of the field strength $\phi$ corresponding to the value $h_{max}$ at four different conformal times: $\eta=1.41 \times 10^{-8}\; \textrm{GeV}^{-8}$ \protect\subref{eta=141,29} and $\eta=2.11 \times 10^{-8}\; \textrm{GeV}^{-1}$ \protect\subref{eta=211,43}, $\eta=2.51 \times 10^{-8}\; \textrm{GeV}^{-1}$ \protect\subref{eta=251,49}, $\eta=3.02 \times 10^{-8}\; \textrm{GeV}^{-1}$\protect\subref{eta=301,55}. Lengths are given in units of the lattice spacing i.e. $10^{-10}\; \textrm{GeV}^{-1}$. \protect\label{network}}
\end{figure}

A~visualization of an exemplary network of Higgs domain walls and its evolution is shown in the figure \ref{network}. The isosurface of the field strength $\phi$ corresponding to the position of the maximum $h_{max}$ approximated by the interpolation from values at lattice sites is visualized for four different conformal times: $\eta=1.41 \times 10^{-8}\; \textrm{GeV}^{-8}=141\, l$ and $\eta=2.11 \times 10^{-8}\; \textrm{GeV}^{-1}=211\, l$, $\eta=2.51 \times 10^{-8}\; \textrm{GeV}^{-1}=251\, l$, $\eta=3.02 \times 10^{-8}\; \textrm{GeV}^{-1}=302\, l$. In the first panel \ref{eta=141,29} we see several domain walls spreading over whole lattice at the beginning of the scaling regime. The further decay of the network during the simulation, presented in panels \ref{eta=211,43} and \ref{eta=251,49}, ends with only one bubble of one vacuum in the background of the second vacuum (panel \ref{eta=301,55}). The visualization was prepared with data from numerical simulations performed on a~lattice with size $256^3$. Percolation theory predicts that networks evolving in the scaling regime consist of walls spreading over the whole Universe. Thus, large sheets visible in the figure \ref{network} shows that a~metastable network can be formed for the specific initial configuration of the field.

Finally, we want to stress that even though this example of the evolution of Higgs domain walls in the scaling regime gives evidence for the occurrence of the generally predicted metastability of networks of domain walls with the potential with nearly degenerate minima, the realisation of this scenario in the early Universe was unlikely. As our numerical simulations show, the formation of the long lived network of Higgs domain walls required not only the Higgs effective potential with nearly degenerate minima, but also the specific initial configuration of the Higgs field. The problem of the production of the metastable network of Higgs domain walls corresponds to simultaneously fine-tuning of two parameters to unnaturally small values. 

\section{Gravitational waves\label{spectrum}}
We use a numerical algorithm proposed in \cite{Dufaux:2007pt} for computation of the GWs produced during inflation and later used also in simulations of domain walls \cite{Kawasaki:2011vv,Hiramatsu:2013qaa,Krajewski:2016vbr}.

The algorithm is based on Einstein equation linearised around FRW metric background solution. We assume that domain walls evolve in the radiation domination epoch during which the scale factor $a(\eta)$ grows linearly:
\begin{gather}
\frac{d a}{d \eta}= a(\eta)^2 H(\eta) = const =: \dot{a}_{in},\\
a(\eta) = \dot{a}_{in} \eta + a_{in}, \label{scale_factor}
\end{gather}
where $H(\eta)$ is the value of the Hubble constant at the conformal time $\eta$ and $a_{in}$ is a~constant which sets the value of the scale factor $a$ at the time when radiation starts to dominate the Universe. We neglect the constant $a_{in}$ in our numerical simulations, assuming that the evolution of domain walls took place deep in the radiation domination era, i.e. the ratio $\frac{a_{in}}{a(\eta)}$ was small at the time when domain walls were formed.

We neglect the back reaction from domain walls to te evolution of the homogeneous gravitational background. This approximation is supported by the fact that the critical energy density in our simulations is much larger than the difference between values of the maximum and the electroweak minimum of the potential $V^\Lambda_{\textrm{SM}}(h_{max}) - V^\Lambda_{\textrm{SM}}(\minEW)$.
We also assume that Higgs field fluctuations were the main source for GWs (constituted the dominant contribution to the transverse-traceless part $T^{TT}$ of the energy-momentum tensor $T$). The lowest order in perturbations of metric $\xi$ around FRW background contribution to $T^{TT}$ is given by
\begin{equation}
T^{TT}_{ij} = \sum_{i,j,l,m} \mathcal{O}_{ijlm} \partial_{x^l} \phi \partial_{x^m} \phi,
\end{equation} 
where $\mathcal{O}_{ijlm}$ are projections on the transverse-traceless part and $\phi$ is the field which models the Higgs field in our simulations.

The linearized Einstein equation can be solved using Green's function and Fourier transform in the background given by eq. \eqref{scale_factor} if one neglects the back reaction on the Higgs field from GWs. The solution used in \cite{Dufaux:2007pt} is of the form
\begin{equation}
\widehat{\xi}_{ij}(\eta,k) = \frac{2}{a(\eta) {M_{Pl}}^2} \int_{\eta_i}^{\eta} d\eta' \frac{\sin \left(|k|\left(\eta - \eta'\right)\right)}{|k|} a(\eta') \widehat{T^{TT}}_{ij} (\eta',k),\label{source_solution}
\end{equation}
where $\widehat{\xi}_{ij}$ is the Fourier transform of $\xi_{ij}$ with the normalization:
\begin{equation}
\widehat{\xi}_{ij}(\eta,k)=\int_{\mathbb{R}^3} d^3 x\ e^{-ikx} \xi_{ij}(\eta,x), \label{Fourier}
\end{equation}
and $\eta_i$ is the conformal time before which the source $\widehat{T^{TT}}_{ij}$ appeared. Moreover, in the Fourier space projection operators can be expressed in a more compact form:
\begin{equation}
\mathcal{O}_{ijlm} (k) := P_{il} (k) P_{jm} (k) - \frac{1}{2} P_{ij} (k) P_{lm} (k),\label{projection_O}
\end{equation}
where $P_{ij}(k) : = \delta_{ij} - \frac{k_i k_j}{|k|^2}$.

The spectrum of gravitational waves' energy density per unit logarithmic frequency interval can be calculated using~\cite{Dufaux:2007pt}:
\begin{equation}
\frac{d \rho_{gw}}{d \log |k|} (\eta,k) = \frac{1}{(2 \pi)^2 {M_{Pl}}^2 {a(\eta)}^4 V} S(\eta,k), \label{energy_density}
\end{equation}
with the function $S$, computed in our simulations on the lattice, given by:
\begin{equation}
\begin{split}
S(\eta,k) =& \frac{|k|^3}{4 \pi} \int_{S^2} d \Omega_k \sum_{i,j,l,m} \left[\left|\int_{\eta_i}^{\eta_f} d\eta' \cos \left(|k|\left(\eta - \eta'\right)\right) a(\eta') \mathcal{O}_{ijlm} \widehat{T}_{lm} (\eta',k)\right|^2 \right.\\
&\left. + \left| \int_{\eta_i}^{\eta_f} d\eta' \sin \left(|k|\left(\eta - \eta'\right)\right) a(\eta') \mathcal{O}_{ijlm} \widehat{T}_{lm} (\eta',k)\right|^2\right],
\end{split}\label{S_expression}
\end{equation}
where $\int_{S^2} d \Omega_k$ denotes the integration over the direction of the wave vector $k$.

In our implementation of the algorithm we calculate six independent arrays of $\partial_{x^i} \phi \partial_{x^j} \phi$ and use the Fourier transform to obtain the energy-momentum tensor $\widehat{T^{TT}}_{ij}$. We calculated the projection operators $\mathcal{O}$ in advance for three families of directions: along edges ($k_i \propto \delta_{1i}, \delta_{2i}, \delta_{3i}$), along diagonals of walls ($k_i \propto \delta_{1i} + \delta_{2i}, \delta_{2i} + \delta_{3i}, \delta_{3i} + \delta_{1i}$) and along the diagonal ($k_i \propto \delta_{1i} + \delta_{2i} + \delta_{3i}$) of the cubic array of momenta. The integration over directions of the wave vector $k$ is substituted by the average of values in distinguished families.

The spectra of GWs obtained from our numerical simulations correspond to a~very early time in the evolution of the Universe when the decay of domains walls ended. In order to compare this data with sensitivities of detectors of GWs we must also calculate how GWs emitted from domain walls were stretched by the expansion of the Universe. Solution \eqref{energy_density} predicts that the energy density of GWs during the radiation domination epoch scales as $a^{-4}$.
Basing on equation \eqref{scale_factor} the ratio of values of the scale factor at the end of the decay of domain walls $a(\eta_{dec})$ to the value at the time of equality of matter and radiation energy densities $a(\eta_{EQ})$ can be expressed by the ratio of proper values of the Hubble constant:
\begin{equation}
\frac{{a(\eta_{dec})}}{{a(\eta_{EQ})}} = \sqrt{\frac{H(\eta_{EQ})}{H(\eta_{dec})}}.
\end{equation}
Values of the Hubble constant at both conformal times can be easily computed. We estimate the value of the Hubble constant at the time of equality of matter and radiation energy densities assuming simple scaling of these densities:
\begin{equation}
{H_{EQ}}^2=\frac{2 {H_0}^2 {\Omega_M}^4}{{\Omega_R}^3}=2\times 10^{-37} {\rm GeV}^2, \label{equlibrium}
\end{equation}
from present day values of the Hubble constant $H_0=1.42\times 10^{-42}{\rm GeV}$ and fractions of the critical density of matter $\Omega_M=0.3$ and radiation $\Omega_R=9.5 \times10^{-5}$. We compute the value of the Hubble constant at the end of the decay of domain walls from parameters of the simulation
\begin{equation}
H_{dec}=\frac{1}{a_{dec}\eta_{dec}} \left(1-\frac{a_{in}}{a_{dec}}\right) =10^{10} \left(1-\frac{a_{in}}{a_{dec}}\right) \left(\frac{10^{-10}\; \textrm{GeV}^{-1}}{a_{dec} \eta_{dec}}\right) \textrm{GeV}. \label{simulation}
\end{equation}
The ratio $\frac{a_{in}}{a_{dec}}$ is relevant for theories in which the radiation domination epoch begins at very low energy scales. In most inflationary scenarios this value is negligible.
Finally, substituting equations \eqref{equlibrium} and \eqref{simulation} we obtain:
\begin{equation}
\frac{{a(\eta_{dec})}}{{a(\eta_{EQ})}} = 7.1\times10^{-24} \left(\frac{10^{10} \textrm{GeV}}{H_{dec}}\right)^{\frac{1}{2}}.
\end{equation}

Furthermore, assuming that the energy density of GWs scales as $a^{-4}$ from the epoch of equality to the present day, we can write:
\begin{equation}
\frac{d \rho_{gw}}{d \log |k|} (\eta_{0},k) = (1+z_{EQ})^{-4} \frac{{a(\eta_{dec})}^4}{{a(\eta_{EQ})}^4} \frac{d \rho_{gw}}{d \log |k|} (\eta_{dec},k),
\end{equation}
where $\eta_{0}$ is the present time and $z_{EQ}$ is the red-shift to the epoch of matter-radiation equality. The energy density of gravitational waves $\frac{d \rho_{gw}}{d \log |k|} (\eta)$ is usually presented as a fraction $\Omega_{gw} (\eta)$ of the critical density $\rho_{cr}(\eta):= {M_{Pl}}^2 H^2(\eta)$.

In order to connect values of the wave vectors on the lattice with the present frequency of GWs we take into account that the wavelength $\lambda(\eta)$ of the GW at any time $\eta$ with the comoving wave vector $k$ at the time of the decay of domain walls satisfy
\begin{equation}
k a(\eta) \lambda (\eta) = 2 \pi, \label{wavelength}
\end{equation}
thus
\begin{equation}
f_0:=f(\eta_0)=\frac{a(\eta_{dec})}{a(\eta_0)} \frac{k}{2 \pi},
\end{equation}
where $\eta_0$ is the present conformal time.
Using the previously estimated value of the red-shift we get:
\begin{equation}
f_{0}=5.07\times 10^{6} \left(\frac{10^{10} \textrm{GeV}}{H_{dec}}\right)^{\frac{1}{2}} \left(\frac{k}{10^{10}\; \textrm{GeV}}\right)\ \textrm{Hz}. \label{redshited_f}
\end{equation}

Using this algorithm in lattice simulations raises some complications. The modified eom \eqref{PRS} with $\alpha=3$ and $\beta=0$ cannot be used, because the modification disturbs dynamics of short wavelength fluctuations.
For the unmodified eom the width of the domain walls decreases as $\propto a^{-2}$.
The width of domain walls at the end of the simulation expressed in units of the lattice spacing cannot be too small in order to correctly model the profile of walls.
On the other hand, the width of domain walls at the initialisation cannot be too large, because many domain walls need to be present on the lattice to properly reproduce the statistical properties of the network. Those requirements significantly restrict the dynamical range of the simulation and consequently only late domain walls ($\eta_{start} = \mathcal{O}(10^{-10}\; \textrm{GeV}^{-1})$) can be reliably investigated in our simulation.

Moreover, as noted in \cite{Hiramatsu:2013qaa}, this algorithm produces a spectrum that diverges as $k^3$ (due to the factor $|k|^3$ in \eqref{S_expression}) for random initialisation of the field strength. Following \cite{Hiramatsu:2013qaa} we have introduced a~cut-off scale of the order of the width of domain walls in the Fourier transform of the initialisation configuration.
However, imposing the cut-off decreases the number of degrees of freedom on the lattice (the long wave-length correlations are induced) and the accuracy of the reproduction of the initialisation probability distribution. Thus, the computation of spectra of GWs produced by networks of domain walls from fine-tuned initial configurations requires accordingly large lattice sizes.

We calculated spectra of GWs produced by decaying networks of Higgs domain walls formed around the conformal time of the order of $\eta_{start} = \mathcal{O}(10^{-10}\; \textrm{GeV}^{-1})$. We used the value of the suppression scale $\Lambda$ which leads to the effective potential with nearly degenerate minima. Our simulations were initialised with configurations of the field generated randomly from probability distributions satisfying $\theta=0$.
We used the value $\sigma=4.75 \times 10^{10}\; \textrm{GeV}$ in the initialisation probability distribution. Used values of $\Lambda$ and $\sigma$ correspond to the most stable late networks observed in our scans over generic initialisation conditions.

\begin{figure}[t]
\centering
\includegraphics[width=0.7\textwidth]{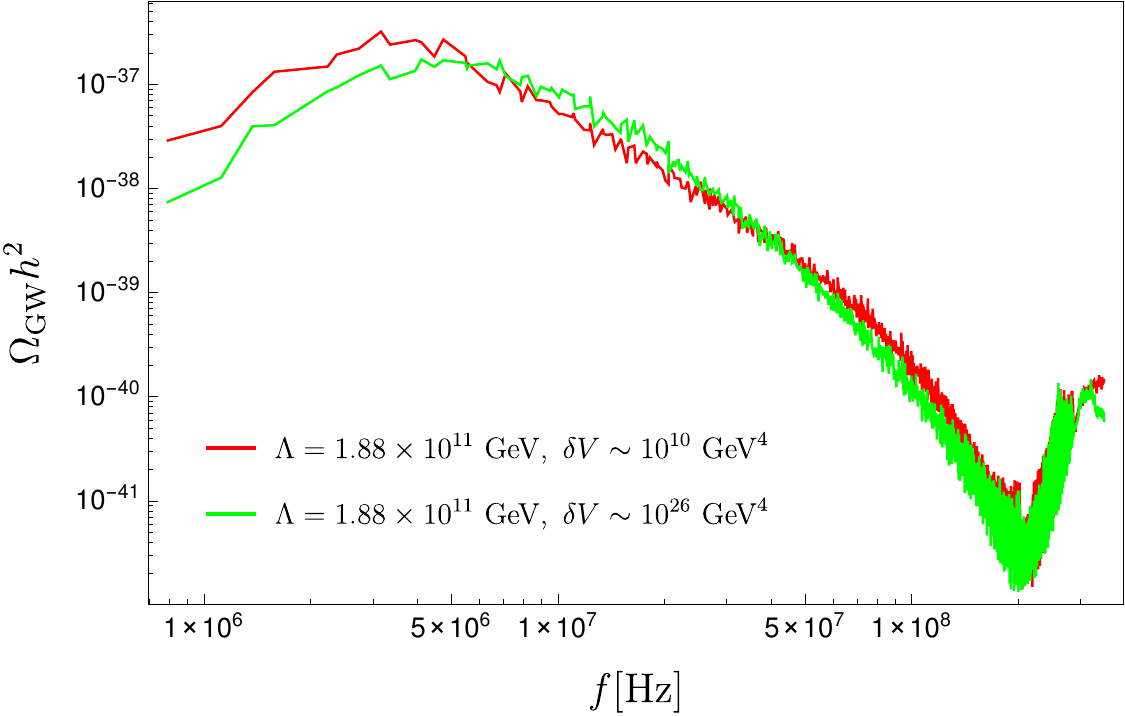}
\caption{Present day spectrum of gravitational waves $\Omega_{gw}$ emitted from Higgs domain walls in the case of the Higgs potential with nearly degenerate minima with the difference in values of potential in minima of the order of $\delta V \sim 10^{10}\; \textrm{GeV}^4$ (red) and $\delta V \sim 10^{26}\; \textrm{GeV}^4$ (green). \protect\label{Omega_GW}}
\end{figure}

The figure~\ref{Omega_GW} shows the present spectra of GWs $\Omega_{gw}(f)$ emitted from Higgs domain walls for two values of $\Lambda$ leading to $\delta V \sim 10^{10}\; \textrm{GeV}^4$ (red) and $\delta V \sim 10^{26}\; \textrm{GeV}^4$ (green). Each of these spectra is the average over five simulations performed on the lattice of the size $512^3$ and initialised at $\eta_{start}=10\, l$ with $a_{start}:=a(\eta_{start}) = \frac{2}{5}$. Simulations ended at the conformal time equal to $\eta_{end}=100\, l$ with $a_{end}:=a(\eta_{end}) = 4$. We also assumed that $a_{in}=0$ and the value of the Hubble constant at the end of simulations is equal to $(a_{end}\eta_{end})^{-1}$. 
The width of domain walls at this time was of the order of $10\, l$. The resulting spectra are peaked at the frequency of the order of \mbox{$f_{peak} \sim 2\times10^{6}\ \textrm{Hz}$} and their amplitude is very small. 

\begin{figure}[t]
\centering
\includegraphics[width=\textwidth]{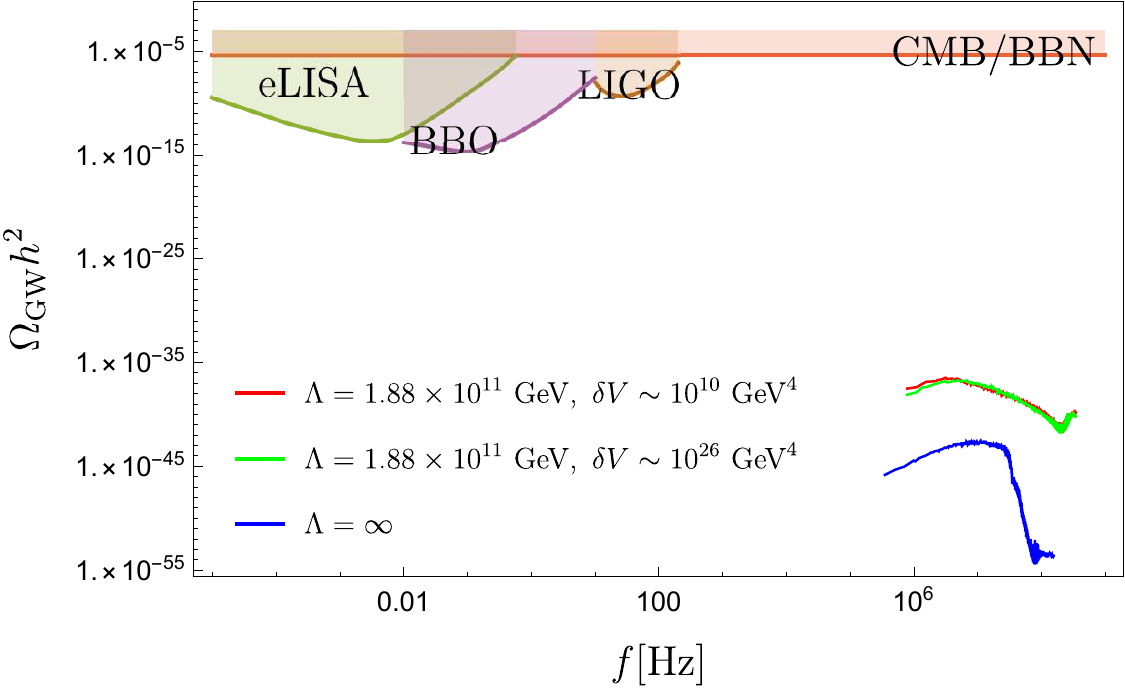}
\caption{Predicted sensitivities (shaded regions) for future GWs detectors: aLIGO, LISA, and BBO compared with the spectrum of GWs (solid) calculated in lattice simulations for the initial values of $\sigma=4.50 \times 10^{10}\; \textrm{GeV}$, $\theta=0$ and the effective potential with nearly degenerate minima with the difference between values of potential in minima of the order of $\delta V \sim 10^{10}\; \textrm{GeV}^4$ (red) and $\delta V \sim 10^{26}\; \textrm{GeV}^4$. The spectum of GWs produced in the SM without nonrenormalizable operators (blue) obtained in \cite{Krajewski:2016vbr} is presented for comparison. \label{sensitivity}}
\end{figure}

The figure \ref{sensitivity} shows the resulting GW spectra (solid) and sensitivity of present and future detectors LIGO \cite{TheLIGOScientific:2014jea}, LISA \cite{Bartolo:2016ami} and BBO \cite{Yagi:2011wg} and the bound on additional energy from CMB/BBN~\cite{Henrot-Versille:2014jua,Smith:2006nka} (shaded regions). Spectrum obtained in our previous work \cite{Krajewski:2016vbr} for Higgs domain walls in the SM without nonrenormalizable operators is also shown in the figure~\ref{sensitivity} for comparison. The spectrum of GWs produced by domain walls in the case of nearly degenerate minima has higher amplitude than the spectrum obtained in the case of the pure SM. The difference is a result of longer life-time of domain walls when nonrenormalizable operators are present. Moreover, the delayed decay of the network effects in the shift of the position of the peak towards lower frequencies. The position of the peak corresponds to the energy scale given by the Hubble radius, thus the later the decay of network of domain walls ends the lower is the frequency of the maximum of the spectrum.

The figure~\ref{sensitivity} shows that even though the energy density emitted in the scenario with nearly degenerate minima of the effective potential is higher than the previously computed value in the case of pure SM, it is still orders of magnitude lower than predicted sensitivity of planned detectors of GWs. This result excludes the possibility of the detection of GWs emitted by SM domain walls in upcoming years for the generic initial configuration of the Higgs field.

\section{Summary\label{summary}}
We investigated the evolution of networks of cosmological domain walls of the Higgs field in the presence of corrections from physics beyond the Standard Model.
We modelled the influence of heavy states not described by the SM, in the framework of the effective field theory by inclusion of the nonrenormalizable operator $h^6$ suppressed by a scale $\Lambda$.
 We have considered a~large range of values of $\Lambda$ starting from the Planck scale and ending at $\Lambda=1.79 \times 10^{11}\; \textrm{GeV}$ at which the high field strength minimum turns into a saddle point. 
We also determined that the electroweak minimum and the high field strength minimum of the RG improved effective potential are degenerate for the value of $\Lambda$ around $1.88 \times 10^{11}\; \textrm{GeV}$.

In section \ref{wall} we described properties of domain walls interpolating between the electroweak minimum of the RG improved effective potential and the second minimum located at the high field strengths.
We determined the width of domain walls to range from $3.5 \times 10^{-9}\; \textrm{GeV}^{-1}$ to $5.0 \times 10^{-9}\ \textrm{GeV}^{-1}$ for the considered range of values of the suppression scale $\Lambda$. In addition we calculated the shift of the position of the local maximum separating two minima induced by the inclusion of the operator $h^6$.

We used numerical lattice simulations based on the PRS algorithm. The same algorithm was used in our previous work \cite{Krajewski:2016vbr} and based on that experience we have run simulations for two different values of the initialisation conformal time: $\eta_{start}= 10^{-12}\; \textrm{GeV}^{-1}$ and $\eta_{start}= 10^{-10}\; \textrm{GeV}^{-1}$.

Networks of cosmological domain walls could be produced in the early Universe by a variety of processes. In this paper we concentrated on the possibility that Higgs field fluctuations interpolating through the potential barrier separating two minima of the potential were generated during cosmological inflation. Inflationary models predict configurations of the Higgs field well approximated by the Gauss distribution with the standard deviation $\sigma_I$ proportional to the Hubble constant during inflation $H_I$. However, the mean value of the Higgs field does not change significantly during inflation so it can not be determined just by an inflationary model.
In this study we simply assumed the most promising, from the point of view of phenomenology, scenario in which the mean value of the probability distribution $\theta$ vanishes. Our numerical simulations prove that configurations with the initial standard deviation lower than $1.2\, h_{max}$ decay to the electroweak vacuum observed today.

Our recent results show that the influence of the nonrenormalizable operator on the evolution of Higgs domain walls was negligible as long as the suppression scale $\Lambda$ was greater than $10^{12}\; \textrm{GeV}$.
For higher $\Lambda$ our previous results obtained in the pure SM \cite{Krajewski:2016vbr} hold. For lower values of $\Lambda$ dynamics of Higgs domain walls change. With the scale $\Lambda$ decreasing down to the value at which two minima are degenerate the maximal allowed (by the requirement of the proper final state) standard deviation $\sigma$ increases due to the shift of the position of the local maximum to higher field strengths. However, the maximal (allowed by the experimental data) ratio $\frac{\sigma}{h_{max}}$ stays constant and approximately equal to $1.2$. Furthermore, late domain walls i.e. formed in simulations initialised at $10^{-10}\; \textrm{GeV}^{-1}$ have longer decay times than early ones (from simulations initialised at earlier conformal times). For values of $\Lambda$ for which the electroweak minimum is the global minimum of the RG improved effective potential, all simulations with $\theta=0$ at the initialisation end with the electroweak vacuum as the final state.

We find that networks of Higgs domain walls produced by generic initialisation conditions are highly unstable. The life-times of such generic networks decaying into the electroweak vacuum predicted by lattice simulations are shorter than $10^{-8}\ \textrm{GeV}^{-1}$.
Moreover, we find no significant increase in the decay time of networks of Higgs domain walls for the effective potential with nearly degenerate minima.

Such short life-times call into question the results of previous studies of networks of domain walls with potentials with nearly degenerate minima in which these networks were found to be metastable. However, majority of previous studies considered approximate global symmetries, leading to potentials symmetric up to small corrections. This is not the case in the SM where the scalar potential is highly asymmetric. Thus, our numerical simulations show that previous results can not be straightforwardly extrapolated to the case of the SM with nonrenormalizable operators. The example of the SM proves that degeneracy of minima of the potential does not necessarily implicate metastability of networks of domain walls. Dynamics of domain walls driven by asymmetric potentials may differ significantly from the one observed with nearly symmetric ones.

Our numerical simulations show that life-times of networks could be longer if their initial properties were more specific.
We considered networks produced by probability distributions satisfying $\theta + \frac{\sigma}{n} = h_{max}$ for $n=300,1000,3000$ and $n=10000$. In these fine-tuned scenarios we observed life-times as long as $5 \times 10^{-8}\; \textrm{GeV}^{-1}$. Initial conditions need to be fine-tuned in order to compensate for asymmetry of the Higgs effective potential around the local maximum. 

Our simulations with fine-tuned conditions prove the possibility of formation of metastable networks of cosmological domain walls when the minima of the potential are nearly degenerate. Moreover, we showed that these long-lived networks of domain walls evolved in the scaling regime satisfying main assumption of semi-analytical methods used in \cite{Kitajima:2015nla}. However, semi-analytical methods do not accommodate for effects of asymmetry of the potential, other than the difference of values between minima. On the other hand, requirement of removing unphysical, short wave-lengths modes from initial configurations limits the usefulness of lattice simulations for computing the energy spectrum of GWs. The calculation for highly fine-tuned initial conditions requires very large lattice size and we were not able to calculate explicitly the spectrum in our lattice simulations in this case

We obtained long-lived networks only for very specific values of the suppression scale $\Lambda$, thus both the coupling constant and the initial configuration of the field need to be fine-tuned. Requirement of the simultaneous fine-tuning of two, a~priori not related, values makes the realisation of this scenario in the nature highly unlikely.

Finally, we computed the energy spectrum of gravitational waves produced during the decay of networks of Higgs domain walls. The long-lived networks were of main interest because the later the process of the decay of domain walls ended the larger the present energy density of produced gravitational waves is predicted to be. We used the value of the suppression scale $\Lambda$ which give the RG improved Higgs effective potential with nearly degenerate minima. 

The spectrum of GWs produced by domain walls in case of nearly degenerate minima has higher amplitude than the one previously obtained in the case of the SM without nonrenromalizable operators \cite{Krajewski:2016vbr}. The difference is the result of longer life-time of domain walls when nonrenormalizable operators are present. Moreover, the delayed decay of the network effects in the shift of the peak towards lower frequencies. Its position corresponds to the Hubble radius, thus the later decay of the network ended, the lower is the frequency of the maximum of the spectrum. Even though, the inclusion of nonrenormalizable operators may lead to enhanced energy density of GWs, the one computed for generic initial configurations is still orders of magnitude to low to be detected in the planned detectors. 

However, the amplitude of the spectrum of GWs can be larger if the evolution of the network of domain walls took place in the specific conditions. As we pointed out previously in \cite{Krajewski:2016vbr}, the present energy density of GWs produced by domain walls could be greater if the evolution of the network of domain walls did not take place deep in the radiation era. Models predicting very low scale of the inflation \cite{Artymowski:2016ikw} or including new components of energy density which shorten the radiation domination period \cite{Lewicki:2016efe,Huang:2016odd} would result in larger $\frac{a_{in}}{a_{dec}}$ (and lower $H_{dec}$). However, this possibility is hard to investigate using lattice simulations due to their small dynamical range and requires semi-analytical extrapolations. 

\acknowledgments{
This work has been supported by the Polish NCN grants DEC-2012/04/A/ST2/00099,\linebreak[4] 2014/13/N/ST2/02712 and 2016/23/N/ST2/03111 and MNiSW grant IP2015 043174 and by the ARC Centre of Excellence for Particle Physics at the Terascale (CoEPP) (CE110001104) and the Centre for the Subatomic Structure of Matter (CSSM).
}

\appendix

\section{Discretisation of equations of motion}
In recent paper we used a~discretisation scheme at first proposed in \cite{Press:1989yh}. It was widely used in the past for numerical simulations of dynamics of domain walls for example in \cite{Lalak:2007rs,Lazanu:2015fua}. Let us consider the following generalisation of eq. \eqref{SM_eom}
\begin{equation}
\frac{\partial^2 \phi}{\partial \eta^2} + \alpha \left(\frac{d \log{a}}{d \log{\eta}}\right) \frac{1}{\eta} \frac{\partial \phi}{\partial \eta} - \Delta \phi + a^\beta \frac{\partial V}{\partial \phi}=0, \label{PRS}
\end{equation}
proposed in \cite{Press:1989yh}. The equation \eqref{PRS} with values $\alpha=3$ and $\beta=0$ combined with used discretisation scheme, proposed in \cite{Press:1989yh}, are known as PRS algorithm. 

Moreover we have used the adaptive time step method which has been proven in the past to be very useful in cosmological lattice simulations \cite{Lalak:2007rs,Krajewski:2016vbr}. More specifically, in our simulations the time step $\Delta\eta$ is calculated from condition that the maximal value (over the lattice) of the relative error generated in the integration step which can be estimated as:
\begin{equation}
\frac{\left(\Phi(\eta+\Delta\eta)-\Phi(\eta)\right)-\left(\phi^{n+1} - \phi^n\right)}{\phi^n} \propto \frac{\ddot{\phi}^{n} - \ddot{\phi}^{n-1}}{\phi^n{\Delta\eta}^2}, \label{relative_error}
\end{equation}
must be smaller than some small constant number $\kappa$. Using this method decreases the number of needed time steps leading to reduction of both: a~computational time and errors coming from finite accuracy of the representation of floating point numbers.

\bibliographystyle{JHEP}
\bibliography{DWSMNO}
\end{document}